\documentclass[useAMS,usenatbib,usegraphicx]{mn2e}

\title[The Open Cluster NGC\,1981]
{Characterization and Photometric Membership of the Open Cluster NGC\,1981}
\author[F. F. S. Maia, W. J. B. Corradi and J. F. C. Santos Jr.]
{F. F. S. Maia\thanks{E-mail: kicage@fisica.ufmg.br},
W. J. B. Corradi, J. F. C. Santos Jr.\\
Instituto de Ci\^{e}ncias Exatas, UFMG, Av. Ant\^{o}nio Carlos 6627, 
Belo Horizonte, Brazil}

\begin{document}

\date{Accepted xxx. Received xxx; in original form 2010 April 09}

\pagerange{\pageref{firstpage}--\pageref{lastpage}} \pubyear{2002}

\maketitle

\label{firstpage}

\begin{abstract}
Open clusters belonging to star-forming complexes are the leftovers
from the initial stellar generations. The study of these young systems provides
constraints to models of star formation and evolution as well as to the
properties of the Galactic disc. We aimed at investigating NGC\,1981, a young 
open cluster in the Orion Nebula Region, using near-IR and $BV(RI)_C$ 
photometric data.
We devised a method that accounts for the field contamination and allows to 
derive photometric membership for the cluster stars. 
A new cluster centre was determined by Gaussian fittings to the 2-D stellar
distribution on the sky, and has been used used to obtain the radial stellar 
density profile and the structural parameters. 
Mass functions were computed
for stars inside the cluster limiting radius and total mass estimated from them.
Although more easily distinguished by its grouping of 6 relatively
bright stars, an underlying population of faint pre-main sequence
stars is evident in the cluster area. We showed that this population is
related to the cluster itself rather than to the nearby Orion Nebula cluster.
Additionally a fraction of the cluster low mass stars may have been 
evaporated from the region in its early evolution leading to the present 
sparse, loose structure.
The estimated parameters of NGC\,1981 are core radius 
$R_\mathrm{c}=0.09\pm0.04$ pc, limiting radius $R_\mathrm{lim}=1.21 \pm 0.11$ 
pc, age $t=5\pm1$ Myr, distance modulus ($m-M$)$_0=7.9\pm0.1$ ($380 \pm 
17$ pc), reddening $E$($B-V$)$=0.07\pm0.03$ and total mass $m = 137 \pm 
14$ M$_\odot$.
\end{abstract}


\section{Introduction}

The open clusters' fundamental physical parameters are important 
pieces of information for 
studies on the formation and evolution of the Galactic disc 
\citep[e.g., ][]{pca95,Jacobson,Sestito} and as grounding 
tests for star formation and evolution models 
\citep[e.g., ][]{lvd96,d02,Siess2000}. 
\citet[hereafter DAML02]{DAML02} have summarized the 
available information on open clusters in a major catalogue 
which continuously expands.  In 
version 2.10\footnote{http://www.astro.iag.usp.br/$\sim$wilton} of this 
catalogue, 1787 open clusters are listed out of which 54 per cent have known 
distance, age and 
reddening, 24 per cent have also proper motion and radial velocity
and only 10 per cent have  metal abundance determined. 
\citet{Kharchenko2005} have created an open 
cluster catalogue with uniformly determined parameters, among them cluster
radius, core radius and age, by means of an 
automated computational algorithm (The Catalogue of Open Cluster Data,
COCD)\footnote{http://www.univie.ac.at/webda/cocd.html}. This catalogue has 
been used to investigate the local ($<$\,1 kpc) population of star cluster 
complexes revealing new constraints on the Galaxy's structure and kinematics
\citep{Piskunov}.

Sky surveys like 2MASS \citep{2mass} produced large
amounts of near-IR data and have contributed to the discovery of even more
objects. Additionally, several studies have been benefited from 2MASS 
database by employing near-IR photometric analyses of structural and 
populational properties of Milky Way star clusters
\citep[e.g.,][]{sbb05,Pavani2007,bsb06}. Because the 2MASS database covers 
the whole sky, it allows data extraction from spatially unlimited regions.
Also, near-IR wavelengths are particularly sensitive to
discriminate cluster stars from the contaminating field for 
young stellar systems \citep[e.g.,][]{sbb05}.

NGC\,1981 (OCl\,525, C\,0532-044) is a young, sparse cluster 
located $\sim$1$^{\circ}$ North from
the Orion Nebula at Galactic coordinates $\ell=208.09^{\circ}$ and 
$b=-18.98^{\circ}$. As noted by \citet{sha52},
\emph{the Orion Nebula seems to be separated from NGC\,1981 only as
a result of a somewhat nearer obscuring cloud.} 
NGC\,1981 depicts the northern end of Orion's Sword.
According to \citet{sgsb95}, NGC\,1981 and Collinder\,70 constitute 
a possible binary cluster being less than 20 pc apart.
The cluster belongs
to Gould's Belt, a planar distribution of O and B stars inclined 
$\sim$\,20$^{\circ}$ with respect to the Galactic plane \citep[][and
references therein]{p01}. In a detailed study of Gould's Belt
\citep{l68}, the brightest 
stars in the cluster (HD\,37016, HD\,37017 and HD\,37040) had their 
spectral types determined (B2.5V, B1.5V, B2.5IV, respectively) and photometric 
distances estimated (all at $494$ pc). 

Subsequent works
emphasized kinematical properties of NGC\,1981 stars, among other clusters, 
aiming at obtaining dynamical and structural characteristics of our Galaxy
\citep[e.g., ][]{h87}. 
More recently, \citet{Kharchenko2005} performed an 
analysis of the cluster as a stellar system
deriving reddening $E$($B-V$)$=0.05$, distance modulus 
($V-M_v$)$=8.16$ ($d=400$ pc), $\log{t({\rm yr})}=7.50$, cluster radius 
$R=0.25^{\circ}$ (1.7 pc) and core radius $R_\mathrm{c}=0.13^{\circ}$ (0.9 pc).
We should note, however, that the age estimated by \citet{Kharchenko2005}
relies upon a single star, as a consequence of the method applied, which
progressively removes stars with low membership 
probabilities from the CMD by means of an iterative process. 

Most studies of the cluster are inserted in more general analyses of
the Orion star formation complex.  It is part of Orion OB1 association,
subgroup c, which is 2 to 6\,Myr old and located closer to us
($\approx$400 pc) than the younger Orion Nebula by at least 10 pc
\citep{Bally2008}. However, such a division in subgroups has been questioned
in favour of a continuous star forming event \citep{Muench}.
Since NGC\,1981 bright stars are early B spectral types, a plausible
evolutionary sequence would entail supernovae explosions from O-type
progenitors causing the compression of the interstellar medium and formation
of the present younger populations.
On the kinematical side, \citet{Dias06} derived 
membership probabilities on the basis of UCAC2 proper motions 
aided by a statistical method for 158 stars within 15 arcmin of 
the cluster centre.

\citet{eac09} investigated hierarchical 
star formation in Gould's Belt based on the spatial and kinematical 
distribution of star clusters, either bound star systems or transient
stellar condensations with a mean lifetime of $\sim$\,10\,Myr or less. 
NGC\,1981 was classified by them as a bound cluster
in view of its age ($\sim$\,30\,Myr) as determined by \citet{Kharchenko2005}. 
In the present work, we argue  that NGC\,1981 is indeed younger, as has
been shown to be the case for Platais\,6 and NGC\,2546, which also belong to
Gould's Belt \citep{eac09}.

Despite the efforts to better understand the astrophysical processes 
occurring in the Orion Nebula and its surroundings and their connection
with Gould's Belt, NGC\,1981 have not been given particular attention 
\citep[except for][]{Kharchenko2005} and, as a consequence, its parameters 
can be improved.
In this work, we present Johnson-Cousins $BV(RI)_C$ photometric observations 
carried out at the Observat\'orio do Pico dos Dias (Itajub\'a, Brazil) along 
with near-IR photometric data extracted from 2MASS survey to study the open 
cluster structural properties and stellar content, aiming at
a better determination of radius, age, distance, reddening and mass.

This work is structured as follows. Sect. 2 describes
the observational data employed. Sect. 3 presents the procedure to obtain
the cluster centre, a fundamental step to build a reliable radial density
profile, discussed in Sect. 4. Colour-magnitude diagrams are employed in
Sect. 5, where
a statistical decontamination algorithm is applied to the cluster CMDs,
and in Sect. 6, where a comparison of data with isochrones allows a
determination of the cluster's astrophysical parameters. In Sect. 7 the 
radial mass function of the cluster is analysed and in Sect. 8 the stellar 
density charts used to map the spatial distribution of stars and their 
frequency in different CMD regions are described.
Further discussions and conclusions are given in Sect. 9.

\section{Data}

\subsection{Near-infrared}

\begin{figure}
\centering
\includegraphics[width=0.95\linewidth]{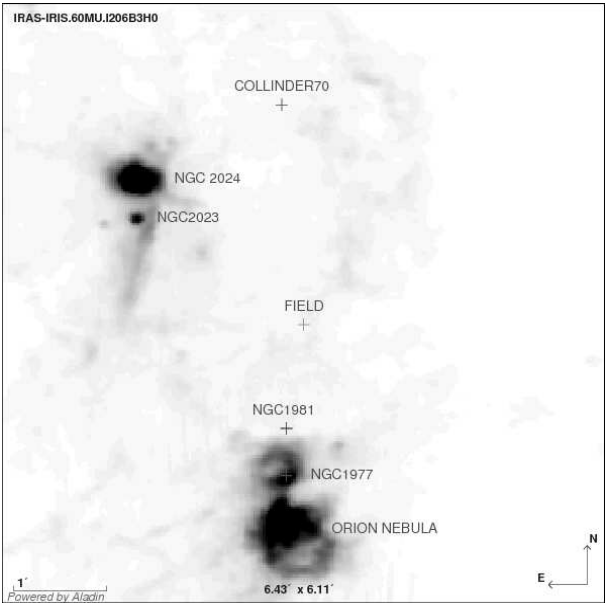}
\caption{NGC\,1981 and the nearby region from IRAS 60\,$\umu$m band. Other 
objects in the area and the selected comparison field are also indicated.}
\label{f:orion}
\end{figure}

Vizier\footnote{http://vizier.u-strasbg.fr/viz-bin/VizieR} was used to extract
near-infrared photometric data from 2MASS in circular fields
centred on equatorial coordinates (J2000) 
$\alpha= 05^\mathrm{h} 35^\mathrm{m} 09^\mathrm{s}$ and 
$\delta= -04^\circ 25\arcmin 54\arcsec$, the cluster centre as 
taken from DAML02. The data encompasses the point sources within $R<55$ arcmin,
corresponding roughly to four times the estimated visual 
radius of the cluster. For decontamination purposes, a comparison field was 
extracted 1$^\circ$ northwest of the cluster, in a region with the same area 
and similar absorption as deduced from near-infrared (2MASS) and mid-infrared 
(IRAS) images. This choice was preferred over an annular field extraction 
centred in the cluster because of the nebulosity associated with NGC\,1977 
towards the south. Figure \ref{f:orion} shows an image of NGC\,1981 and the 
nearby region from IRAS 60\,$\umu$m band. The selected comparison field and 
other objects are also shown. 

As photometric quality constraint, the 2MASS extracted data were
restricted to unambiguous point sources brighter than specified on the 99 per 
cent Point Source Catalogue completeness limit 
(15.8, 15.1 and 14.3 mag for 10-$\sigma$ at $J$, $H$, and $K_s$, respectively).

\subsection{Optical}

CCD $BV$($RI$)$_C$ observations were carried out with the 0.6-m (IAG) telescope
at the Observat\'orio do Pico dos Dias  (Laborat\'orio Nacional de 
Astrof\'{\i}sica, Itajub\'a, Brazil) during the night of September 21, 2000. 
The CCD detector employed was a $1$$\times$$1$k SITe SI003AB (\#\,101) 
configured to yield readout noise 5.5\,e$^-$ and gain 5.0\,e$^-$ADU$^{-1}$. 
The plate scale was 1.22 arcsec.pixel$^{-1}$. The images were obtained using a 
focal reducer to cover a field-of-view of 21 $\times$ 21 arcmin and 
were taken with a mosaic setup on 6 adjacent positions around the cluster 
centre in order to cover the full extension of the object. 
To transform instrumental magnitudes into the standard system, 15 objects from 
\citet{Landolt} standard stars catalogue were also observed on the same night, 
with 2 standard stars being used for atmospheric extinction correction. Details
of the observations are shown in Table \ref{t:obslog}, where the observed 
fields are listed along with their exposure times, typical seeing values and 
airmasses.

\begin{table}
\centering
\caption{Observation log for NGC\,1981}
\begin{tabular}{lcccc}
\hline
\hline
Cluster/Region & Filter & Exposure (s)& Seeing (${\arcmin \arcmin}$)& Airmass \\
\hline
                   & B  & 1 15 50  & 3.4 & 1.2 \\
 NGC\,1981/CE      & V  & 1 20  & 2.5  & 1.2 \\
                   & R  & 1 20 & 2.4 & 1.2 \\
                   & I  & 1 20 & 3.9 & 1.2 \\ \hline
                   & B  & 1 15 50 & 3.1 & 1.1 \\
 NGC\,1981/CW      & V  & 1 20 & 2.8 & 1.1 \\
                   & R  & 1 20 & 2.7 & 1.1 \\ 
                   & I  & 1 20 & 3.4 & 1.1 \\\hline
                   & B  & 1 15 50 & 2.7 & 1.3 \\
 NGC\,1981/NE      & V  & 1 20 & 2.6 & 1.3 \\
                   & R  & 1 20 & 2.3 & 1.2 \\
                   & I  & 1 20 & 3.8 & 1.2 \\ \hline
                   & B  & 1 15 50 & 3.1 & 1.1 \\
 NGC\,1981/N       & V  & 1 20 & 2.6 & 1.1 \\
                   & R  & 1 20 & 2.6 & 1.1 \\
                   & I  & 1 20 & 3.7 & 1.1 \\ \hline
                   & B  & 1 15 50 & 3.3 & 1.1 \\
 NGC\,1981/NW      & V  & 1 20 & 2.9 & 1.1 \\
                   & R  & 1 20 & 2.8 & 1.1 \\
                   & I  & 1 20 & 4.1 & 1.1 \\ \hline
                   & B  & 1 15 50 & 3.3 & 1.1 \\
 NGC\,1981/S       & V  & 1 20 & 2.6 & 1.1 \\
                   & R  & 1 20 & 2.7 & 1.1 \\
                   & I  & 1 20 & 3.7 & 1.1 \\ 
\hline
\end{tabular}
\label{t:obslog}
\end{table}

The data were reduced using the standard {\sc iraf} routines for aperture 
photometry. Besides the standard reduction 
steps, the 6 adjacent fields have been aligned and recurrent stars were 
selected according to their distance to the image centre in each filter. 
Specifically, for every recurrent star, we kept only the measurement of the 
star closest to the image centre. This choice avoided the {\it vignetting} 
effect introduced on the images borders by the focal reducer. The calibration 
equations used to transform to the standard system were:

\begin{eqnarray}
b &=& B + b_1 + b_2X_{b} + b_3(B-V) + b_4(B-V)X_{b} \\
v &=& V + v_1 + v_2X_{v} + v_3(B-V) + v_4(B-V)X_{v} \\ 
r &=& R + r_1 + r_2X_{r} + r_3(V-R) + r_4(V-R)X_{r} \\ 
i &=& I + i_1 + i_2X_{i} + i_3(V-I) + i_4(V-I)X_{i}
\end{eqnarray} 

\noindent
where $B$, $V$, $R$ and $I$ are the standard magnitudes and $b$, $v$, $r$ and
$i$ are the instrumental magnitudes; $X_n$ is the airmass on each filter. The
subscript coefficients $i=1...4$, were obtained from a 2-pass interactive 
fitting on each filter with the extinction coefficients (subscripts 2 and 4) 
being fitted first and then used as constants on a second fit of the zero 
point and colour terms (subscripts 1 and 3 respectively). The derived 
coefficients and the RMS deviation of the fittings to the catalogued magnitudes
of the standard stars are shown in Table \ref{t:coef}.

\begin{table}
\centering
\caption{Coefficients and RMS deviation of the calibration fit.}
\begin{tabular}{c r r r r}
\hline \hline
$i$ & \multicolumn{1}{c}{B} & \multicolumn{1}{c}{V} & \multicolumn{1}{c}{R} & \multicolumn{1}{c}{I} \\\hline
1 & $5.79\pm$0.01  & $5.35\pm$0.01  & $4.53\pm$0.02  & $5.05\pm$0.01\\
2 & $0.28\pm$0.06  & $0.17\pm$0.04  & $0.11\pm$0.04  & $0.08\pm$0.05\\
3 & $-0.19\pm$0.01 & $-0.20\pm$0.02 & $-0.18\pm$0.07 & $-0.11\pm$0.02\\
4 & $-0.04\pm$0.04 & $0.15\pm$0.02  & $0.19\pm$0.04  & $0.07\pm$0.03\\
\hline
RMS & \multicolumn{1}{c}{0.03} & \multicolumn{1}{c}{0.02} & \multicolumn{1}{c}{0.04} & \multicolumn{1}{c}{0.03} \\
\hline
\end{tabular}
\label{t:coef}
\end{table}

As photometric quality constraint the employed data set was that restricted to 
an error less than
$0.04$ mag, corresponding to the higher deviations found on the calibration 
fit. This value implies a magnitude limit of approximately $15$ mag and leads 
to an average magnitude uncertainty of $0.05$ mag (taking into account the 
calibration deviation).

The following analysis involving spatial information were developed using 2MASS
data due to its large extraction area, good astrometry and the poor seeing of 
the optical data, which was employed mostly for CMD comparisons with the 
near-IR data.
                         
\section{New Centre Determination}

\begin{figure}
\centering
\includegraphics[width=\linewidth]{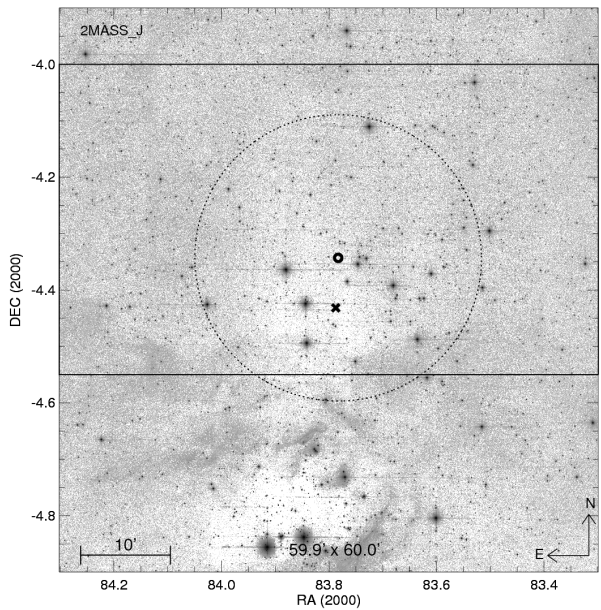}
\caption{Sky chart of NGC\,1981 showing the region used for centre 
determination (rectangle), the literature centre (cross), the 
calculated centre (circle) and the data extraction area for the 
decontamination method (dotted circle).}
\label{f:radec}
\end{figure}

The sparse aspect of NGC\,1981 makes difficult to estimate its centre,
which is an essential step before a reasonable radial density profile can be
obtained. Catalogued centre values are mainly intended for identification 
purposes and are generally imprecise. Therefore, we have estimated the centre
of NGC\,1981 by first selecting a region (see Fig.\ref{f:radec})
around the coordinates given by DAML02. 
The search area was limited to the sky around the catalogued centre and 
did not include stars southward of $\delta =-4.55^\circ$  to avoid the 
nebulosity and contamination from NGC\,1977.

\begin{figure}
\centering
\includegraphics[width=\linewidth]{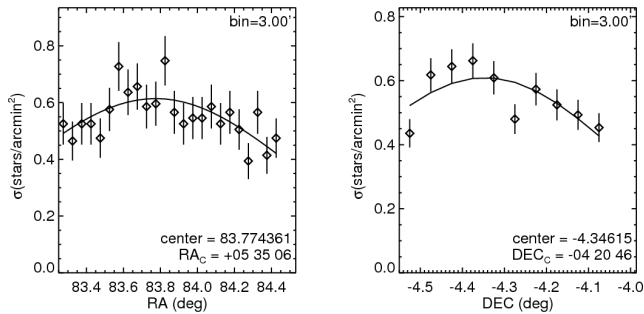}
\caption{Gaussian fittings for the 
determination of central coordinates. Bin size on
RA and DEC is $3.0$ arcmin. Error bars correspond to Poisson 
statistical errors.}
\label{f:cgauss}
\end{figure}

Furthermore, the selected region was divided into bins of right ascension
and declination and star counts were made inside them. We used these star 
counts to build spatial profiles and fitted a Gaussian function to the stars 
distribution on RA and DEC, adopting the centre of the fitted Gaussian as the
cluster centre coordinate. Fig. \ref{f:cgauss} shows the Gaussian fitting 
obtained for bin size $3.0$ arcmin and the resulting coordinates of the 
centre.

\begin{figure}
\centering
\includegraphics[width=\linewidth]{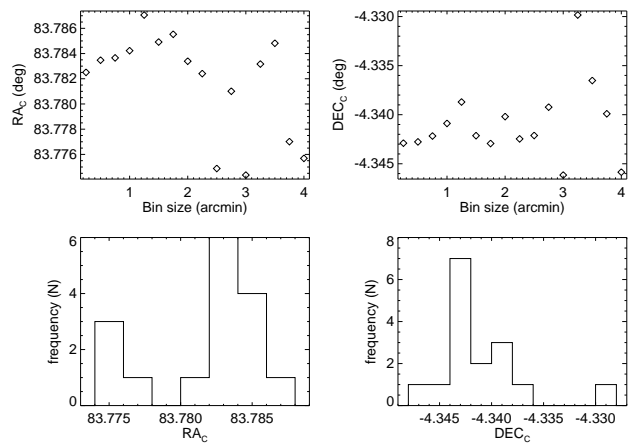}
\caption{Relation between 
determined coordinates and the bin size
(top). Histograms showing the most recurrent right ascension and declination 
for the cluster centre (bottom).}
\label{f:coordbin}
\end{figure}

This procedure was applied to different bin sizes ranging from 
$0.25-4.00$ arcmin and the resulting centre coordinates were used to create 
histograms showing the most recurrent RA and DEC values. 
We adopted these values as the coordinates of the cluster centre. We also 
found no trends between the coordinates and the bin size. Figure 
\ref{f:coordbin} shows this relationship and the histograms defining the 
cluster centre.

The adopted coordinates of the cluster centre are $\alpha~=~83^\circ 46\arcmin
59\arcsec = 5^\mathrm{h} 35^\mathrm{m} 08^\mathrm{s}$ and 
$\delta = -04^\circ 20\arcmin 35\arcsec$. The histograms bin 
width were chosen based on the standard deviation of the centre coordinates and
accounts for an uncertainty of approximately $5$ arcsec on $\alpha$ and 
$\delta$. Fig. \ref{f:radec} 
shows that the calculated centre is $5.32$ arcmin north and $0.27$ arcmin 
east from the literature centre. The area used for centre determination and the
data extraction region for the decontamination method (see Sect.\,5) are also 
shown.

\section{Radial density profile}

The structural properties of star clusters can be derived 
by means of a projected density profile such as the radial density profile 
(RDP). The RDPs of star clusters usually follow an analytical profile and can 
be described by functions with different parameters related to the cluster 
structure. These parameters include the tidal radius, core radius, core stellar
density and background star density. However, young and sparse open clusters 
like NGC\,1981 usually show little overdensity in comparison to the field 
making difficult the fitting of analytical functions and so the determination 
of their structural parameters.

In order to circumvent this problem the density profile was built by 
superposing RDPs constructed by counting stars inside successive radial rings
of fixed width up to 55 arcmin and then dividing by the area of the rings. 
Ring widths ranged from $0.75-2.00$ arcmin, with $0.25$ arcmin increments.
The narrower rings are ideal to probe the core structure of the cluster while 
the larger ones are better to probe the external regions. This prevents either 
region of the cluster to be undersampled. 

\begin{figure}
\centering
\includegraphics[width=\linewidth]{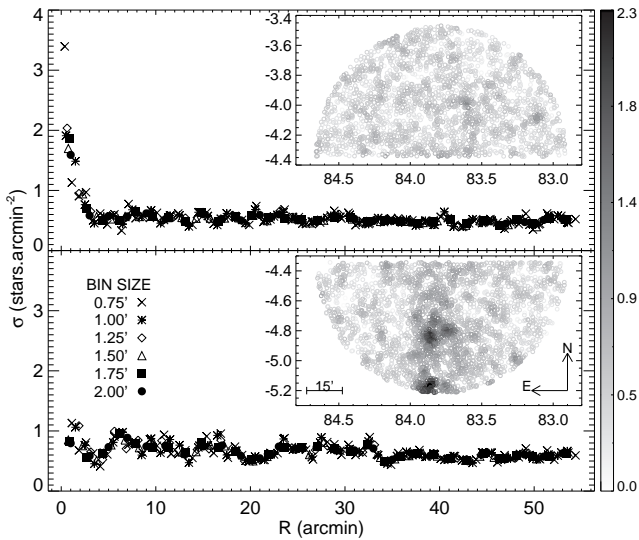}
\caption{RDP of the north (top) and south (bottom) regions of NGC\,1981. 
Stellar density maps of each region are also shown (insets). Radial ring sizes 
range from 0.75$-$2.00 arcmin. Colourbar on the right represents the stellar 
density (stars.arcmin$^{-2}$) on the insets.}
\label{f:northsouth}
\end{figure}

The possible contamination from NGC\,1977 stars to the cluster population was
investigated with separate radial profiles for the northern and southern 
region of NGC\,1981 using semi-circles based on the central declination 
($\delta = -04^\circ 20\arcmin 35\arcsec$) determined in Sect.\,3. 
Density maps were also built up to the extreme radius of $55$ arcmin to 
allow for the discrimination of subtle stellar density differences between  
both regions. To obtain a good contrast on the resulting density map
an optimal radial bin of $2.5$ arcmin was used to calculate the stellar 
density around each star.  Fig. \ref{f:northsouth}
shows the radial profiles of the north (top) and south (bottom) regions and the 
corresponding density maps (insets).

Although NGC\,1977 is about 30 arcmin southwards it can be seen in Fig. 
\ref{f:northsouth} that the contribution from its stars clearly affects the 
density profile of the southern region of NGC\,1981. Since it is not possible 
to tell the boundaries between the two clusters or readily distinguish their 
stars, we turned to the northern part of the density profile to investigate the
structural properties of NGC\,1981 by means of analytical density functions. 

Trials of King profile fittings were 
made on the density profile using the two-parameter 
modified density function introduced by \citet{King1962}:
\begin{equation}
f_2(R) = \sigma_\mathrm{bg} + \frac{\sigma_0}{1+(R/R_\mathrm{c})^2}\ . 
\label{e:kng2} 
\end{equation}

\begin{figure}
\centering
\includegraphics[width=\linewidth]{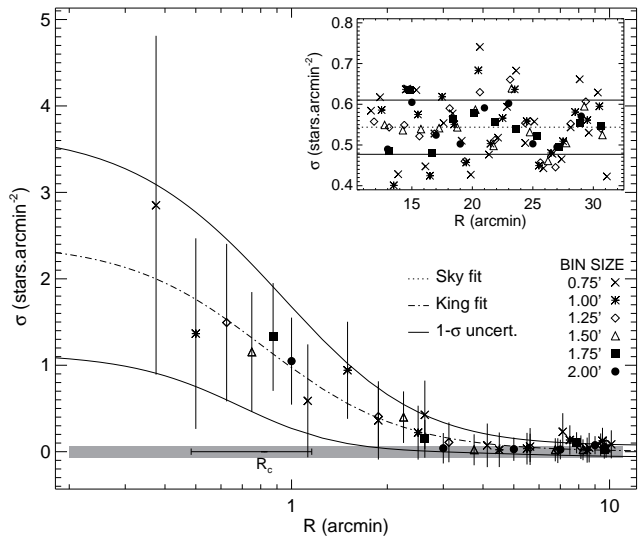}
\caption{2-parameter King-profile fittings. 
Background level was determined beyond 16 arcmin (inset). The determined core 
radius is indicated. Sky fluctuation is represented by a grey rectangle. 
Error bars denote 1-$\sigma$ Poissonian fluctuations.}
\label{f:kngfit}
\end{figure}

To provide better convergence the sky level ($\sigma_\mathrm{bg}$) 
was estimated as the mean of the radial bins inside the range 16-32 arcmin and 
further subtracted from each radial ring to derive the central density 
($\sigma_0$) and core radius ($R_\mathrm{c}$) through the 2-parameter King 
function. The fitting was weighted by the statistical Poissonian errors of 
the star counts inside each radial ring. 
Fig. \ref{f:kngfit} shows the fitting of the 2-parameter function and
the sky determination (inset) for the 
northern region of the cluster with the 1-$\sigma$ uncertainties of the 
fittings. The determined parameters $R_\mathrm{c}$ and $\sigma_\mathrm{bg}$ are
also indicated with their associated errors. 

The determined structural parameters were central density \mbox{$\sigma_0=2.4
\pm1.2$} stars.arcmin$^{-2}$, core radius \mbox{$R_\mathrm{c}=0.83\pm0.34$} 
arcmin and background level \mbox{$\sigma_\mathrm{bg}=0.54\pm0.07$} 
stars.arcmin$^{-2}$. 
A cluster limiting radius ($R_\mathrm{lim}$) was estimated by visually 
inspecting the radius where stars from the cluster are completely merged with 
the background. We found $R_\mathrm{lim}=11\pm1$ arcmin for NGC\,1981. Using 
the distance of $380$ pc (see Sect. 6), $1$ arcmin corresponds to $0.11$ pc.

\section{Colour-magnitude diagrams and field-star decontamination}

Colour-magnitude diagrams of open clusters usually present well-defined 
stellar sequences such as the main sequence, turn-off and giant branch that 
provide essential information on the physical properties of these objects. 
However, the characterization of open clusters is often hindered by strong 
reddening and field-star contamination, specially for the objects projected 
against the Galactic centre or sparsely populated. 

The identification of the cluster stars as opposed to the field-star 
contamination is very important in the study of these objects. 
Assessing membership using proper motion data is 
only available for a small subset of the known open clusters, particularly the 
closest ones. 
An alternative method for disentangling cluster and field-stars consists of 
using photometric data by means of statistical comparison of star samples 
taken from the cluster region and from an offset-field. Our decontamination 
procedure is one of these methods and was based on the work of 
\citet{bb07}.

\subsection{Decontamination procedure}

A field-star decontamination procedure was carried out by firstly selecting an 
appropriate comparison field presenting stellar density and reddening similar
to the cluster. While circular annular fields circumscribed to the cluster area
are appropriated for clusters in dense regions and homogeneous backgrounds, 
they fail on regions affected by differential reddening or rapidly changing 
backgrounds. Therefore, an adjacent field with similar reddening was selected 
for NGC\,1981 considering the nebulosity 
contamination from NGC\,1977 in the southern region of the cluster. Regarding
the cluster area, the data included stars within the inner 16 arcmin of the 
determined centre. This central region corresponds to the cluster 
visual radius, chosen to encompass the brightest B type stars in the field.

We built 3D colour-magnitude diagrams for both cluster and field-stars
with $J$, $J-H$ and $J-K_s$ as axes. These colours provide better 
discrimination among the clusters sequences on the CMD \citep{bbg04}.
The diagrams were divided into small cells of average sizes
$\Delta J=0.6$\,mag, $\Delta(J-H)=0.3$\,mag and $\Delta(J-K)=0.3$\, mag, 
corresponding roughly to ten times the average uncertainties in the colours.
These cells are small enough to detect local variations of field-star 
contamination on the various sequences in the CMD, but large enough to 
accommodate a significant number of stars.
The grids in Fig. \ref{f:dcmd} illustrate these cell sizes. 

Initial cluster membership was assigned to cluster stars within each 
cell based on their overdensity with relation to the field-stars, according to 
the relation 
\mbox{$P=(N_\mathrm{clu}-N_\mathrm{fld})/N_\mathrm{clu}$}. Null 
probability was assigned whenever an excess of field-stars over cluster stars 
occurred in a given cell. 
A subset of the original cluster sample was created by removing from each
cell on the cluster CMD, the expected number of field-stars as measured in 
the control field CMD based on their distance to the calculated centre of the 
cluster. Particularly, each cell on the cluster CMD had the stars 
most distant from the cluster centre removed and cells without cluster 
overdensity had all stars inside their limits removed. 

In order to account for the initial choice of parameters, we applied the 
described method for different grid specifications by changing the position and
size of the cells in each of the CMD axes. Cell positions were changed 
by shifting the entire grid one third of the cell size in each direction. 
Cell sizes were increased and decreased by one third of the 
average sizes in each of the CMD axes. Considering all possible configurations, 
729 different grid sets were used to derive final membership probabilities by 
taking the average, median and mode of the membership obtained in each 
configuration for each star. An exclusion index was also created by noting how 
many times each star was removed from the sample, and then normalizing by the 
number of grid configurations. 

Possible effects of the offset-field selection were investigated by applying 
the method to 3 additional comparison fields 1$^\circ$ distant from the cluster
in different directions northward of the cluster. 
The average, median and mode of the membership of each star, obtained from the 
multiple grid configurations, were compared once different offset-fields were 
employed in the decontamination method. The offset-field selection implied mean 
deviations of 7 per cent on the average membership, 8 per cent on the median 
membership and 
12 per cent on the mode membership. These values were adopted as the general 
uncertainty of these statistical indicators.  
Final membership probabilities were thus
assigned to each star by taking the average of the membership from each grid 
configuration.

The decontaminated sample was obtained by removing stars based on another
two complementary criteria. First, stars with average membership value lower 
than 7 per cent or median membership value lower than 8 per cent were removed. 
Second, stars 
with exclusion index larger than 50 per cent were removed. 
The first criterion ensures the removal of stars that could present null 
membership, based on the uncertainties of each statistical indicator. 
The second criterion is based on the exclusion index and decontaminate the 
cluster sequences by removing the expected field population. The exclusion
index threshold (50 per cent) was set to provide the removal of approximately 
the same number of stars removed by the first criterion.  

\begin{figure}
\centering
\includegraphics[width=\linewidth]{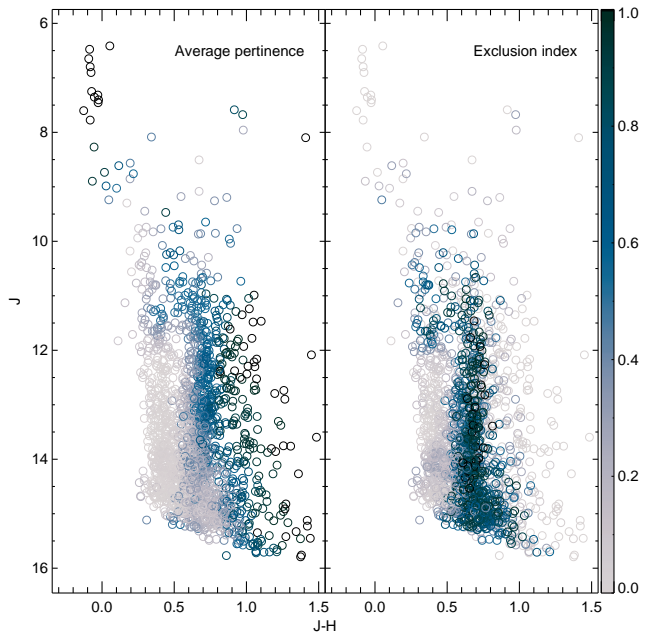}
\caption{Decontamination domain of the complementary adopted criteria. Note 
that stars with lower membership probability ({\it left}) were removed by the 
first criterion whereas stars with higher exclusion index ({\it right}) were 
removed by the second criterion. 
The colourbar indicates both the membership probability (left panel) and 
the exclusion index (right panel) of stars.}
\label{f:decont}
\end{figure}

The adopted criteria complement each other well, acting on different parts of 
the CMD and presenting very little overlapping of the removed stars. The first 
criterion removes stars with low membership probability, acting mainly on 
regions where the cluster and field population are entangled, presenting 
roughly the same density on the CMD. For NGC\,1981 it represents the lower and 
mid-main sequence/pre-main sequence.
For the second criterion, it is worth recalling that stars where removed from 
each cluster CMD cell according to the counts from the offset-field CMD and on 
the distance from the cluster centre. Therefore the second criterion clean the 
cluster sequences from the stars in the outer regions of the cluster and 
effectively removes from the CMD the stars were field population is dominant.
It decontaminates the region of the pre-main sequence stars, mainly for the 
fainter regions ($J > 10$). Fig. \ref{f:decont} shows the decontamination 
domain of each criterion. 

\subsection{Results on NGC\,1981}

\begin{figure*}
\centering
\includegraphics[width=0.6\linewidth,angle=90]{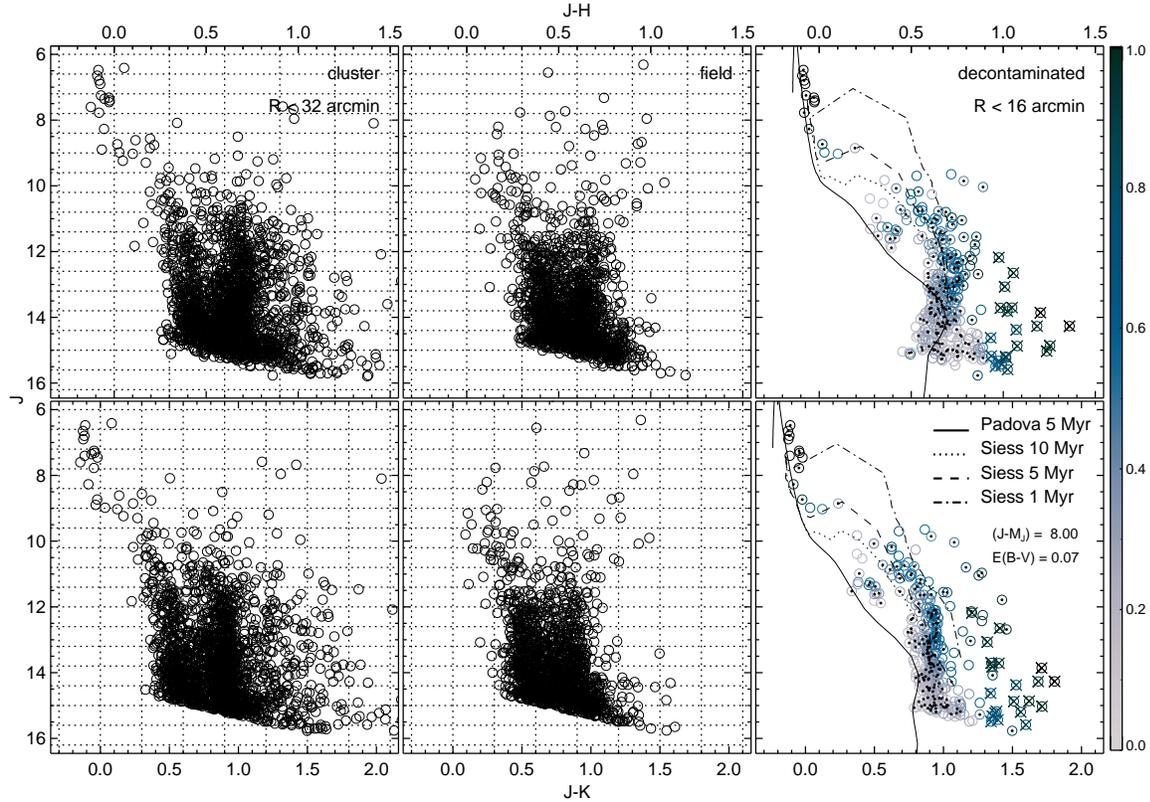}
\caption{Cluster (left), field (middle) and decontaminated (right) CMDs using 
$J-H$ (top) and $J-K_s$ (bottom) colours. Very reddened stars, present in the 
southern region of the cluster were marked with crosses. Stars inside the 
$R_\mathrm{lim}$ are marked with dots. 
The average cell sizes are indicated by dotted grids.
MS and PMS isochrone fittings are also shown on the right panels.
Colourbar on the right indicate the calculated cluster membership probability.} 
\label{f:dcmd}
\end{figure*}

\begin{figure}
\centering
\hskip -0.5cm
\includegraphics[width=\linewidth]{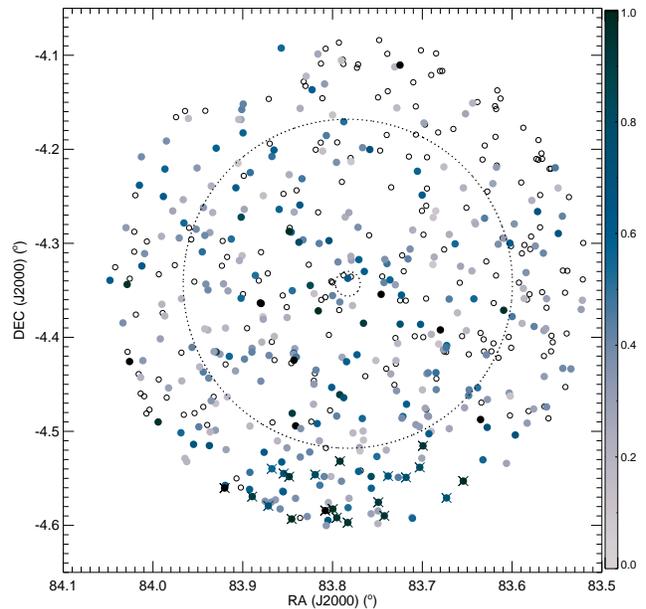}
\caption{Sky chart comparing the removed field population (open circles) and 
remaining cluster stars (filled circles). Concentric circles represent the 
core radius and limiting radius respectively. Very reddened stars are marked 
with crosses. Colourbar indicates assigned membership probabilities.}
\label{f:dchart}
\end{figure}

In order to increase the statistical sample and better probe the field 
population, the procedure was applied on stars within 32 arcmin from the 
cluster centre and the resulting decontaminated sample, used in the subsequent
analysis, was truncated at the visual radius of the cluster (16 arcmin). 
This truncation was done because the adjacent southern clouds associated to 
NGC\,1977 yield a very reddened population which occupy a well defined region
in the CMD (see Sect. 8). The CMD locus of this reddened population does not 
overimpose that
of most cluster stars therefore not interfering with the decontamination
procedure. However, even 
inside the visual radius, very reddened stars presenting $J-H > 0.9$ and 
$J-K_s > 1.2$ appear on the southern region of the cluster (crosses in Fig. 
\ref{f:dcmd}). They were not 
properly sampled by the field population and therefore could neither be 
excluded by the procedure nor present reliable membership probabilities. 
These stars are likely
members of NGC\,1977. Fig. \ref{f:dchart} shows the sky chart of NGC\,1981 
up to 16 arcmin, comparing the member and field population. Structural 
parameters derived in Sect. 4 are also indicated.

The number of stars removed from the contaminated sample do not account for 
the entire field population because the expected number of field-stars measured
on the offset-field is still higher than the number of excluded stars. 
On average, each criterion removed 45 per cent of the field population while 
presenting a common exclusion rate of 5 per cent and a total decontamination 
efficiency of 85 per cent. Most of the exclusion takes place in the crowded 
areas of the CMD as 95 per cent of the stars excluded belong to the 
lower main-sequence ($J > 12$).

The choice of the offset-field accounts for an average deviation of 5 per cent 
in the number of stars left in the decontaminated subsamples (member stars).
Moreover the fraction of recurring member stars, independent of the 
offset-field, is greater than 87 per cent. Fig. \ref{f:dcmd} compares cluster, 
field and decontaminated stars in the $J\,\times\,J-H$ and $J\,\times\,J-K_s$ 
diagrams.

\subsection{Comparison with proper motion data}

Proper motion data from UCAC2 \citep{ucac2} and UCAC3 \citep{ucac3} 
catalogues were used to investigate the reliability
of the photometric memberships determined for the cluster stars.
We selected stars within 15 arcmin from the cluster centre, 
with 2MASS photometry subject to the constraints described in Sect. 2.1 and 
only those with proper motions and their uncertainties.
The corresponding samples amount to 159 
objects in UCAC2 catalogue and 99 objects in UCAC3 catalogue.

\begin{figure}
\centering
\includegraphics[width=\linewidth]{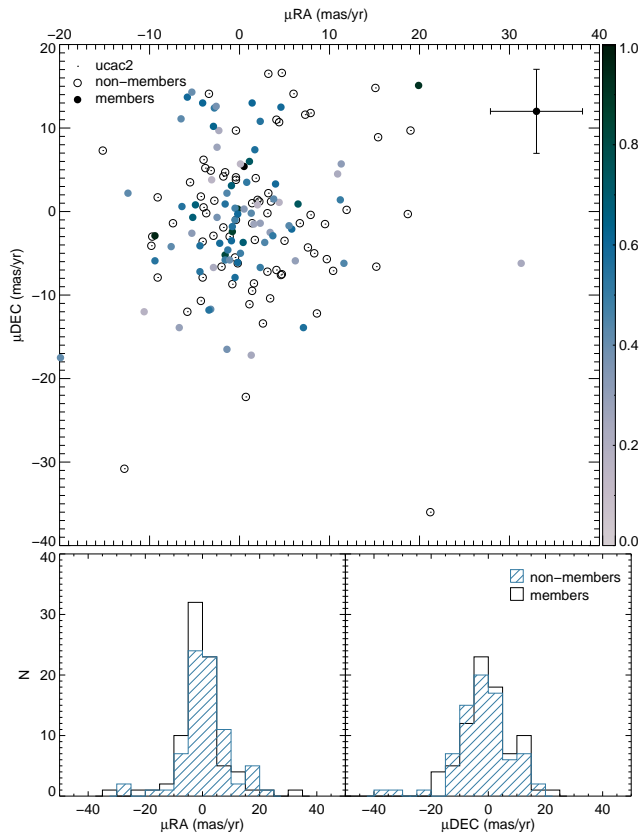}
\caption{Correlation between our membership assignment and proper motion 
data taken from UCAC2 (top). Mean proper motion uncertainties are indicated. 
Probability distribution function of member and non-member stars are also shown in the bottom panels for RA (left) and DEC (right) proper motion directions. }
\label{f:ucac2}
\end{figure}

Comparison of the UCAC2 proper motion with our membership results was done 
using the 2MASS designation to identify the common stars between the datasets.
This allowed us to use our photometric membership to discriminate cluster and 
field-stars on the vector point diagram (VPD) (Fig. \ref{f:ucac2}. top 
panel). Although the photometric members present
a concentration near $\mathrm{(\umu RA,\umu DEC)} = (0,0)$, this is also true 
for the field stars, as can be more easily seen in the probability distribution 
histograms (bottom panels).

As the statistical distance between the cluster and field proper motion
distributions is small, it is difficult to separate the populations.
This effect becomes more accentuated for 
clusters where there is a large field to cluster member ratio or when the 
centroids of the proper motion distributions are very close to each other, such
as in the case of NGC\,1981. 
This problem has already been discussed by \citet{Cabrera90} and 
\citet{Sanchez09} which proposed to use the spatial distribution of the stars, 
in addition to their proper motion distribuition, to increase the statistical 
distance between the populations.

\begin{figure}
\centering
\includegraphics[width=\linewidth]{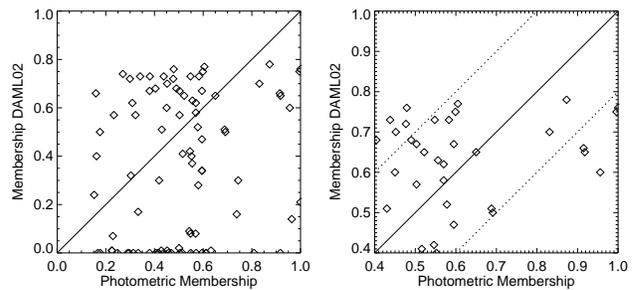}
\caption{Correlation between the proper motion membership from DALM02 and 
our photometric memberships (left). Stars between the dotted lines 
have membership differences smaller than 20 per cent (right).}
\label{f:corpm}
\end{figure}

In fact, membership probabilities have been estimated for NGC\,1981 by 
\citet{Dias06} using UCAC2 proper motion of 160 stars within 15 arcmin of the 
cluster centre. The comparison between these memberships and the photometric 
memberships derived in this work shows a very poor correlation, as can be seen 
in Fig. \ref{f:corpm} (left panel). 
Although the small statistical distance between the populations may account for
the large scattering in the proper motion membership, especially for the 
low-membership stars, it seems that the derived photometric memberships are,
in average, underestimated in relation to the proper motion ones. This trend is
still visible when the low-membership stars are removed from both the 
photometric and proper motion sample, although the correspondence between the 
samples improves as most of the remaining stars present membership differences 
smaller than 20 per cent (right panel). 

\begin{figure}
\centering
\includegraphics[width=\linewidth]{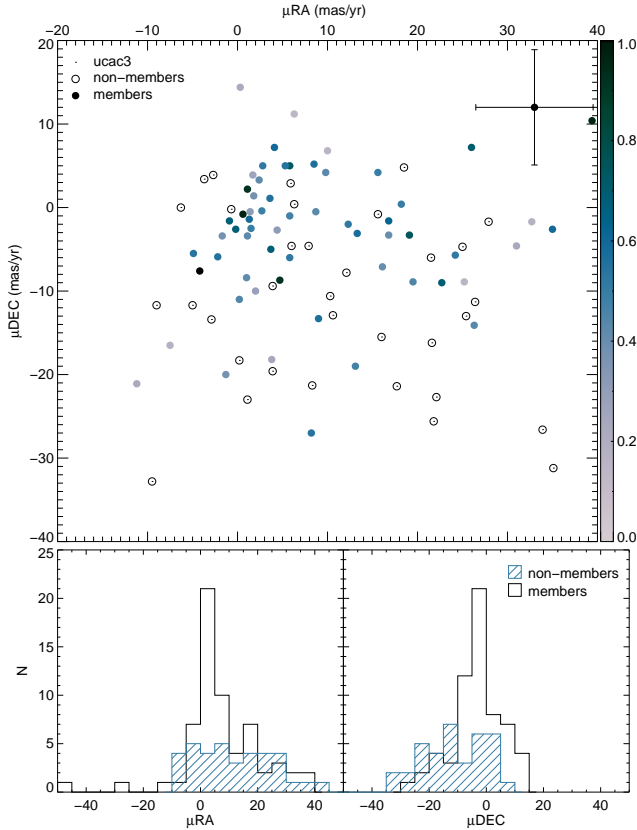}
\caption{Similar to Fig. \ref{f:ucac2}, but using UCAC3 proper motions.}
\label{f:ucac3}
\end{figure}

UCAC3 provided a complete new reduction of the Southern Proper Motion
data that improved the proper motions of faint stars by a factor of 2 compared 
to UCAC2. Although this first release still presents some unsolved problems 
as described in \citet{ucac3}, we present a comparison of our results with 
proper motion from UCAC3 in Fig. \ref{f:ucac3}.
It is clear that UCAC3 VPD is sparser and less populated than that using UCAC2 
data. However, it is currently not possible to realize the origin of this lack
of stars \citep[see Sect. 8 in][]{ucac3}. 
Additionally, while our photometric analysis yields very entangled cluster and 
field populations in UCAC2 VPD, they are much better discriminated
when proper motions from UCAC3 are considered (bottom panels of Fig.
\ref{f:ucac2} and \ref{f:ucac3}). Therefore, membership derived from 
UCAC3 proper motions might be in closer agreement with our photometric 
analysis.

\section{Astrophysical parameters}

The optical data was correlated with the infrared data allowing the membership 
probabilities derived from 2MASS data to be assigned to the stars with $BVRI$
photometry.
Although shallower than the infrared sample and presenting larger photometric 
uncertainties, the optical data provided a larger spectral base for the CMD 
analysis and additional constraints on the physical parameters.

Astrophysical parameters were determined by means of fittings of Padova 
isochrones for 2MASS \citep{bbg04} and 
Johnson-Cousins \citep{Marigo08} filters on the decontaminated CMDs.
Pre-main sequence evolutionary tracks \citep{Siess2000} were also employed. 
We chose only isochrones with overshooting and solar metallicity.
As initial values for the cluster 
physical parameters we were guided by the previous studies of 
\citet{Kharchenko2005} and \citet{Bally2008}.

\subsection{Age}

\begin{figure}
\centering
\includegraphics[width=\linewidth]{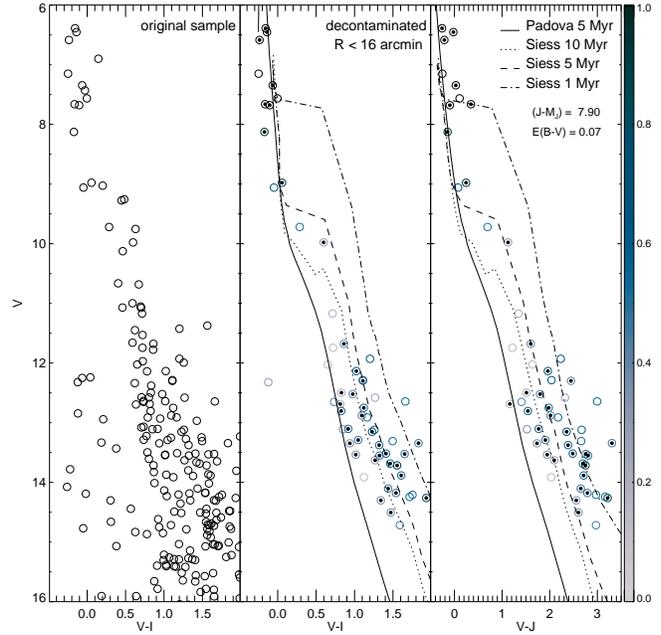}
\caption{NGC\,1981 decontaminated CMDs with optical data. Colours and symbols 
are the same as in Fig. \ref{f:dcmd}.}
\label{f:ocmd}
\end{figure}

The absence of evolved sequences in the CMDs due to the cluster youth makes
determination of its age unreliable if one uses standard evolutionary
models, i.e., those without accounting for pre-main sequence stars.
Pre-main sequence isochrone fittings suggest that stars with ages ranging from 
1-10 Myr coexists in the cluster and that most high-membership ($P > 0.6$)
stars seems to fall between the young pre-main sequence isochrones of 1-5 
Myr. This behaviour is also observed when only the innermost stars (inside the 
limiting radius) are considered. 
Fig. \ref{f:dcmd} shows the isochrone fittings on the decontaminated 
$J \times (J-H)$ and $J \times (J-K_s)$ CMDs along with the derived membership
of stars. Fig. \ref{f:ocmd} shows the same fittings on the $V \times (V-I)$ and
$V \times (V-J)$ CMDs. The average age derived for the cluster is $5\pm1$ Myr
based on isochrone fittings to both near-infrared and optical data.

In photometric studies of star clusters the MS turn-off is the most reliable 
feature for determining the age of these objects. However the identification 
of this CMD feature in very young clusters is uncertain by the scarcity of
massive stars and the isochrones lack of turn-off sensitivity.
For these objects the turn-on, the CMD locus where the PMS stars joins
the MS, is a good indicator of the age of the stellar population.

From the isochrone fittings shown in Fig. \ref{f:dcmd} it 
can be seen that NGC\,1981 turn-on is at $J \approx 9$, 
corresponding to an age of 5 Myr. 
\citet{Cignoni2010} devised a method to derive the age of young stellar 
systems by detecting the PMS turn-on with the luminosity function (LF). The
clustering of stars at magnitudes near the turn-on produces a bump in the LF 
that becomes increasingly fainter as the cluster ages. By using 
synthetic simple stellar populations  they calibrated a relation between the
magnitude $M_V$ of this luminosity bump and the age of the population.  

\begin{figure}
\centering
\includegraphics[width=\linewidth]{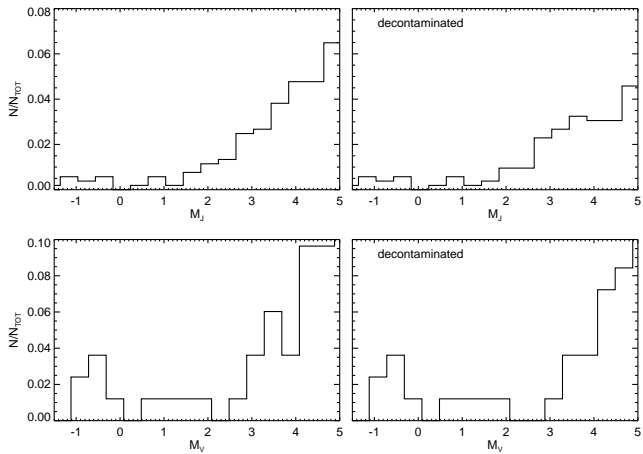}
\caption{Luminosity functions in the infrared (top) and optical (bottom) bands. 
The turn-on bumps at $M_V$=1.3 and $M_J$=0.8 can be identified on 
both original (left) and decontaminated (right) samples.}
\label{f:lf}
\end{figure}

The reddening and distance modulus determined in the present work were used to 
construct LFs with $M_J$ and $M_V$ for both the original and 
decontaminated sample, which are shown in Fig. \ref{f:lf}. In both cases it 
was possible to identify a small bump in the LF at $M_J=0.8\pm0.2$ and 
$M_V=1.3\pm0.6$. 
This bump on the LF correspond to the age $3.3\pm2.7$ Myr, according to the 
age calibration with $M_V$ by \citet{Cignoni2010}.

However, as stated by \citet{Cignoni2010}, reliable 
identification of the turn-on can only be made in populous clusters containing
at least $\approx$ 50 stars brighter than $M_V = 5$ in order to successfully 
detect the bump among the Poissonian fluctuations. Therefore, although we 
should use this result with caution, it is in agreement, within the 
uncertainties, with the age and distance modulus obtained by means of isochrone
fittings.

\subsection{Distance and Reddening}

Although there were small discrepancies in the derived parameters (mainly on 
reddening) from optical versus infrared data, the mean determined values of 
distance modulus \mbox{$J-M_J =8.0 \pm 0.1$} and colour excess 
\mbox{$E(B-V)=0.07 \pm 0.03$} provided good isochrone fitting in both cases. 
With those values the true distance modulus is ($m-M$)$_0=7.9\pm0.1$ where 
we have used $A_J/A_V=0.282$ and $A_V/E(B-V) = 3.09$ \citep{Rieke85}, resulting
in the distance of $380\pm17$ pc.
The uncertainties were set to accommodate the values derived from the optimal 
fittings on both optical and infrared CMDs.  These values 
are similar to the values found by \citet{Kharchenko2005}.
The almost 50 per cent deviation between optical and infrared reddening values 
deserves further discussion.

Although the literature reddening value provided a good isochrone fitting 
on the infrared data, the reddening value obtained from the 
$V \times (V-I)$ CMD suggests that the extinction in the optical bands is 
larger by a factor of 2. 

\begin{figure}
\centering
\includegraphics[width=\linewidth]{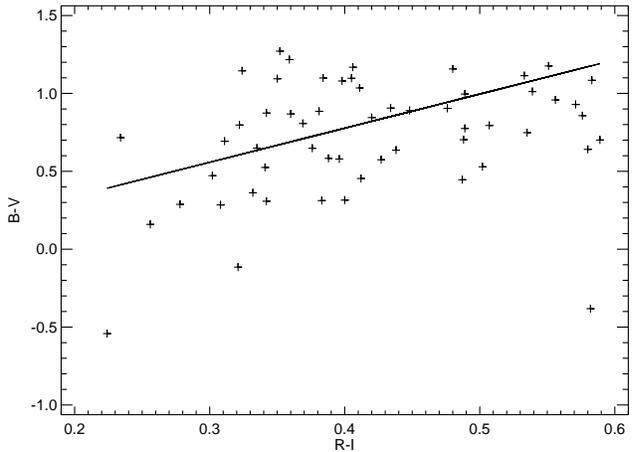}
\caption{Colour-colour diagram with the optical data showing the ZAMS.}
\label{f:redd}
\end{figure}

Reddening in the optical bands was also determined in a previous work 
\citep{Maia} by a linear fit to the 
ZAMS \citep{SK} in colour-colour diagrams as 
proposed by \citet{Munari96}. By selecting 
the diagram most sensitive to reddening (Fig. \ref{f:redd}) and limiting our 
analysis to stars brighter than $V=13$ we derived a reddening value of 
$E(B-V)=0.11$, which is very close to the value obtained from our best 
isochrone fitting on the optical data.

The reddening in the cluster direction, as interpolated in the dust
maps by \citet{sfd98}, is $E$($B-V$)$=0.24\pm0.04$. Since these maps are 
sensitive to dust column density, this value can be considered as an upper 
limit to the cluster reddening. Also the cluster may be affected by 
differential reddening caused by the dust clouds in the region, specially for 
its southern part.

\section{Radial Mass Function}

\begin{figure}
\centering
\includegraphics[width=\linewidth]{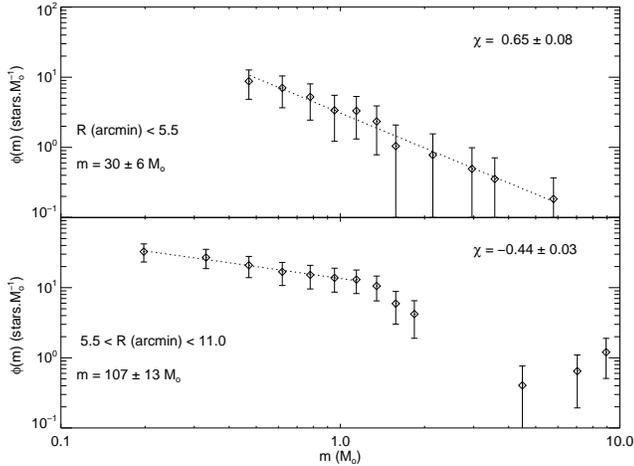}
\caption{Mass function of stars inside the inner 5.5 arcmin (top) and 
within the range $5.5$-$11$ arcmin (bottom). Power law fittings, 
resulting slopes and total mass are indicated. Error bars 
correspond to 1-$\sigma$ fluctuations}
\label{f:mf}
\end{figure}

To analyse the spatial dependence of the mass function (MF), the 
decontaminated LF in the $J$ band determined in Sect. 6.2 
was used to investigate the regions comprehended within the inner 
5.5 arcmin (half limiting radius) and within the range between 5.5 arcmin 
and the limiting radius (11 arcmin) of NGC\,1981. 
The decontaminated LFs were 
converted into one MF for each region by fitting a mass-luminosity relation 
(in $J$ band) from the combined PMS and evolved isochrones of 5 Myr at solar 
metallicity. The power law $\phi(m) = A \cdot m^{-(1+\chi)}$, where $A$ is a 
normalization factor and $\chi$ is the MF slope, was fitted to the data in both 
regions. The MFs were normalized by the number of stars ($N$) inside each 
region ($N=25$ for the inner region and $N=81$ for the outer region).
The outer region MF fit was limited to stars less massive than
$\approx 1$ M$_\odot$ due to its small number of brighter stars.  
The MFs of the selected regions are presented in Fig. \ref{f:mf}, together 
with the power-law fits and the determined slopes.

Both regions have MFs that are flatter than the \citet{Salpeter} one, with 
slopes $\chi = 0.65\pm 0.08$ (inner region) and $\chi= -0.44\pm 0.03$ (outer 
region). 
Total mass inside each region was calculated by summing over selected mass 
ranges. We obtained for the inner region $m = 30 \pm 6$ M$_\odot$ 
($0.4 < m($M$_\odot) < 6$) and for the outer region $m = 107 \pm 13$ M$_\odot$
($0.2 < m($M$_\odot) < 9$). For the inner region the mass was also calculated 
by integrating the derived power law over the same mass range yielding 
$m = 38 \pm 2$ M$_\odot$. We should note that this mass implies an
uniform distribution of stellar masses while the observed distribution of 
the brightest stars suffers from small number statistics. 

The relaxation time of a star system can be defined as 
$t_\mathrm{relax} = \frac{N}{8\ln N}t_\mathrm{cross}$, where 
$t_\mathrm{cross} = R/\sigma_v$ is the crossing time, $N$ 
is the total number of stars and $\sigma_v$ is the velocity dispersion 
\citep{Binney,Lada03}. 
By using the dispersion in proper motion given by DAML02 for 31 stars within 
11 arcmin from the cluster centre with membership probability (derived by 
DAML02) above 50 per cent and assuming
an isotropic distribution of spatial velocities we estimated a mean velocity 
dispersion $\sigma_v \approx 8$ km.s$^{-1}$. Combining this information with 
the limiting radius we get $t_\mathrm{cross}= R/\sigma_v = 0.15$ Myr and 
$t_\mathrm{relax} = 0.41$ Myr for $N= 106$ stars.

Although unexpected 
for such a young object, dynamical evolution can be responsible for rapid 
collapse of the cluster core causing mass segregation and evaporation of stars 
in outer regions. This may have occurred in NGC\,1981 since the relaxation time 
of the cluster is much smaller than its age. 
As shown in Fig. \ref{f:mf} the mass function slope is larger for the inner 
5.5 arcmin than for the outer 5.5-11 arcmin annular region between 
the mass range 0.4-1.1 M$_\odot$. This is in apparent contradiction with the 
expected consequences of dynamical evolution, i.e., the existence of an excess
of low mass stars and depletion of massive ones (eventually grouped in binaries
and/or multiple systems) in the outer region compared to the inner region. 
However 
we indeed do not found any star with mass lower than 0.4 M$_\odot$ in the inner
region but 23 stars in the same mass range in the outer region, in agreement 
with the mass segregation scenario.

\section{Star density maps}

\begin{figure}
\centering
\hskip -0.5cm
\includegraphics[width=0.5\linewidth]{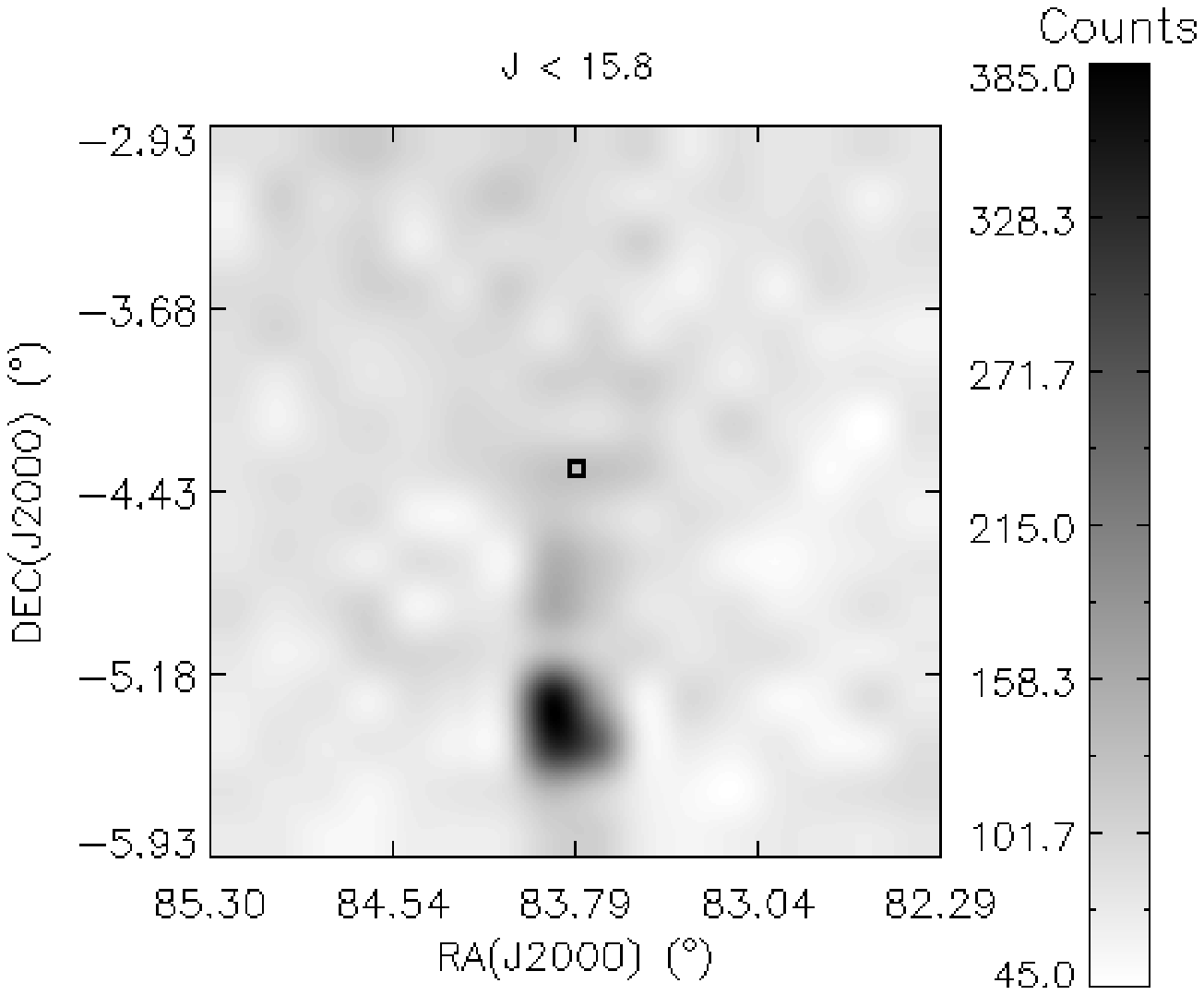}
\includegraphics[width=0.5\linewidth]{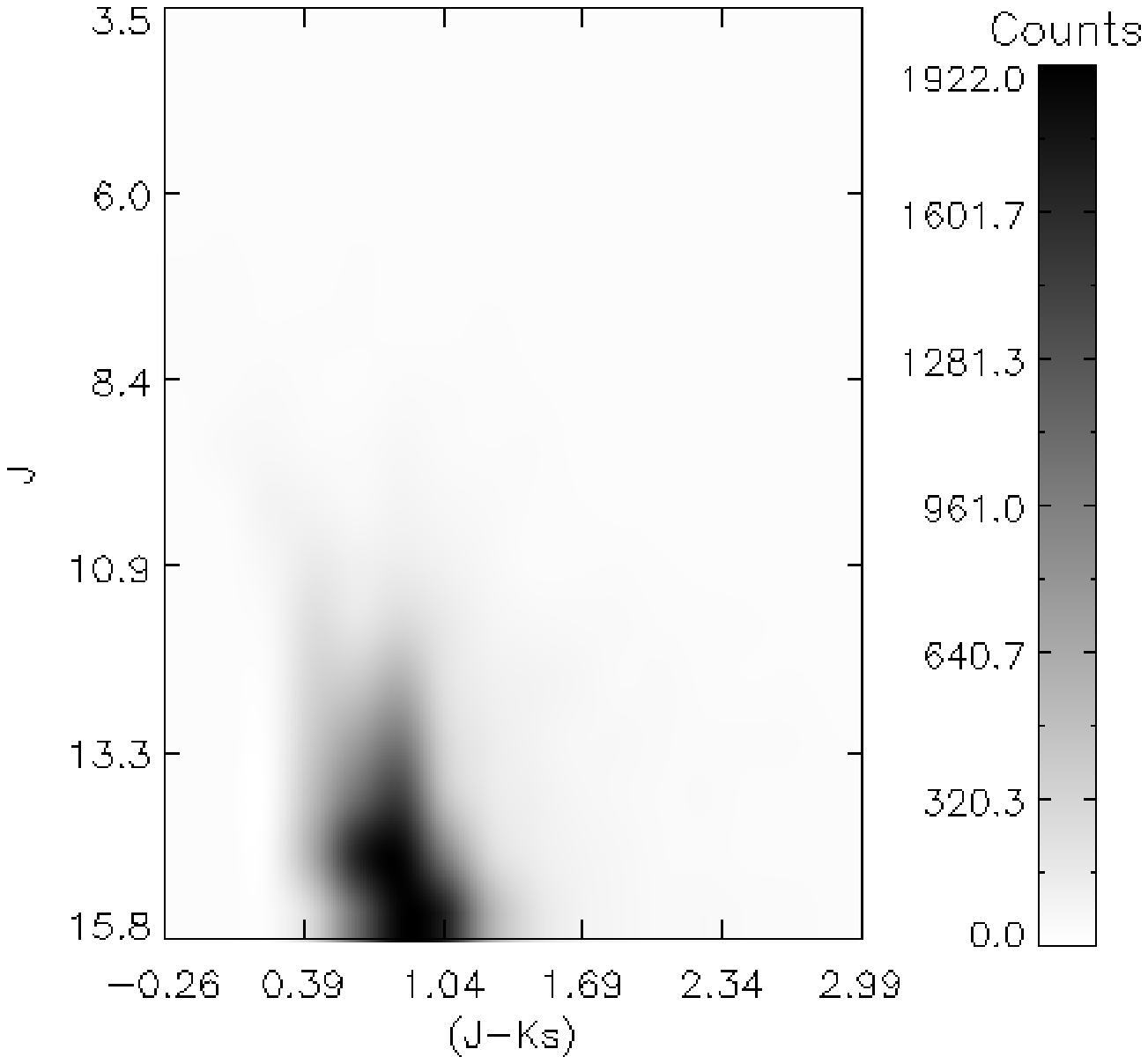}
\caption{Star density map (left) and Hess diagram (right)
of a region 3$^{\circ}$$\times$3$^{\circ}$ centred in NGC\,1981 (square)
based on 2MASS data.}
\label{f:densall}
\end{figure}

We have mapped the projected spatial star density and the frequency
of stars in different positions of the CMD (Hess diagram). Firstly
we show in Fig. \ref{f:densall} the star density map and the corresponding
Hess diagram of a large region around NGC\,1981 including the Orion
Nebula Cluster (ONC). Both diagrams were built by dividing the
plots in square cells of 32 pixels on a side, which in Fig. \ref{f:densall}
corresponds to $5.6$ arcmin for the spatial map, $0.1$ in ($J-K_s$) and 
$0.4$ mag in $J$ for the Hess diagram.
The density diagrams are then smoothed by a cubic convolution
interpolation method.

The data comprises all sources extracted from 2MASS catalogue obeying
the photometric quality constraints mentioned in Sect. 2.1.
The ONC dominates the density map southwards of NGC\,1981 which the centre,
as determined in Sect. 3, is indicated by a square. Even though, an
overdensity compared to the adjacent neighbourhood can be noticed at
the position of NGC\,1981. The Hess diagram provide evidence that most
of the stellar content in the ONC consists of red low mass stars.

\begin{figure}
\centering
\hskip -0.5cm
\includegraphics[width=0.5\linewidth]{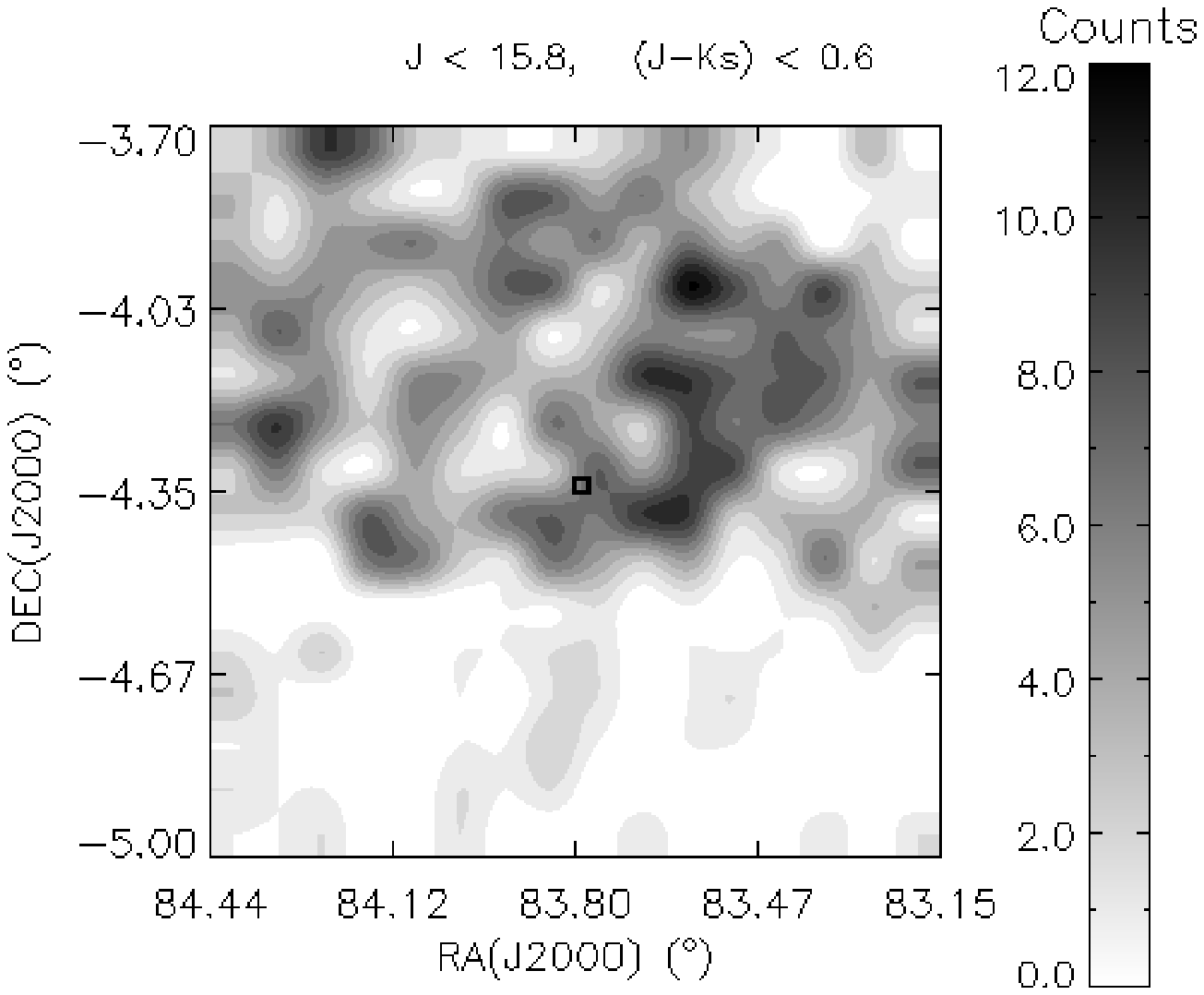}
\includegraphics[width=0.5\linewidth]{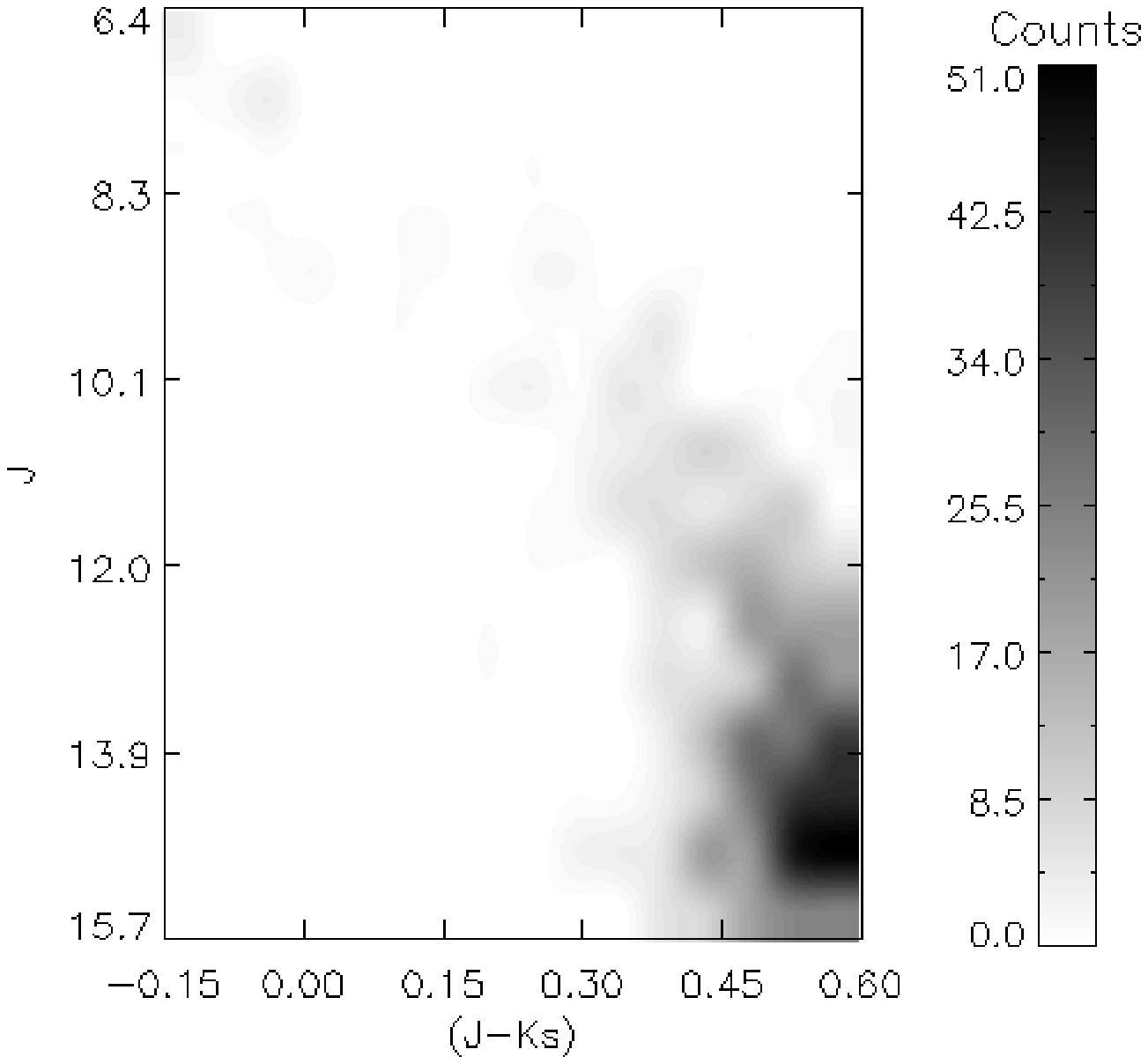}
\caption{Same as Fig. \ref{f:densall} but for a smaller region
centred in NGC\,1981 and ($J-K_s$)$<0.6$.}
\label{f:denspart}
\end{figure}

Smaller sky and CMD areas were considered around NGC\,1981
in Fig. \ref{f:denspart}.
The square cells of 32 pixels correspond to
$2.4$ arcmin for the spatial map and $0.02$ in ($J-K_s$), $0.3$ mag in $J$
for the Hess diagram.
The spatial map has $1.3^{\circ} \times 1.3^{\circ}$ and the
associated Hess diagram
has been limited to contain only stars with $(J-K_s)< 0.6$, thus excluding
the reddest stars. Such a colour constraint clearly separates the spatial map
in two regions: a denser star field northwards of NGC\,1981 (but including it)
and a more rarefied star field southwards of NGC\,1981. It clearly reflects
the presence of dust revealing a boundary between a heavy obscured stellar
field and a region with lower extinction. It also indicates where the
reddest stellar population lies in the area. The cut in colour eliminates
reddened stars by dust but also pre-main sequence stars, known to be
abundant in the ONC region.

\begin{figure}
\centering
\hskip -0.5cm
\includegraphics[width=0.5\linewidth]{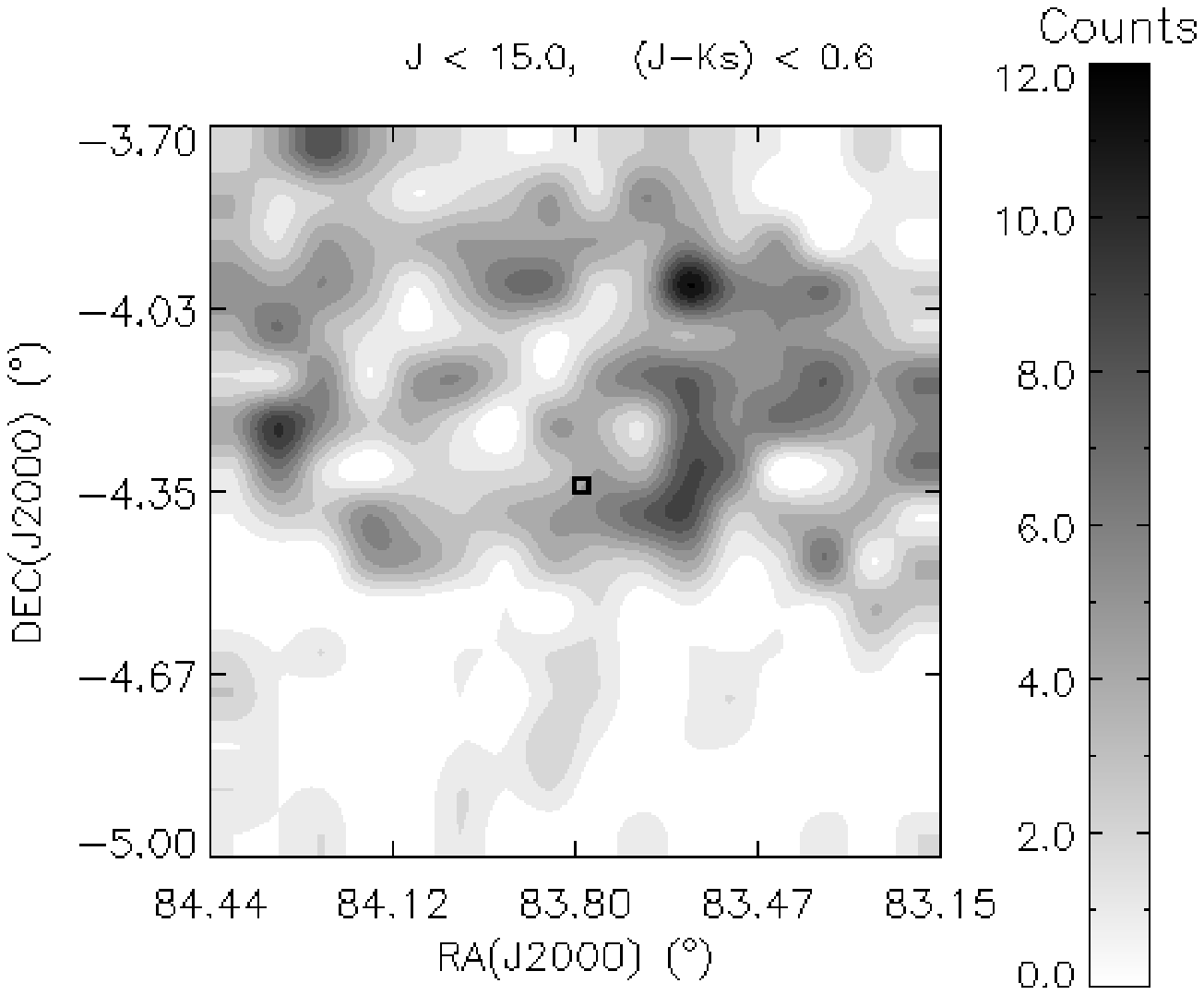}
\includegraphics[width=0.5\linewidth]{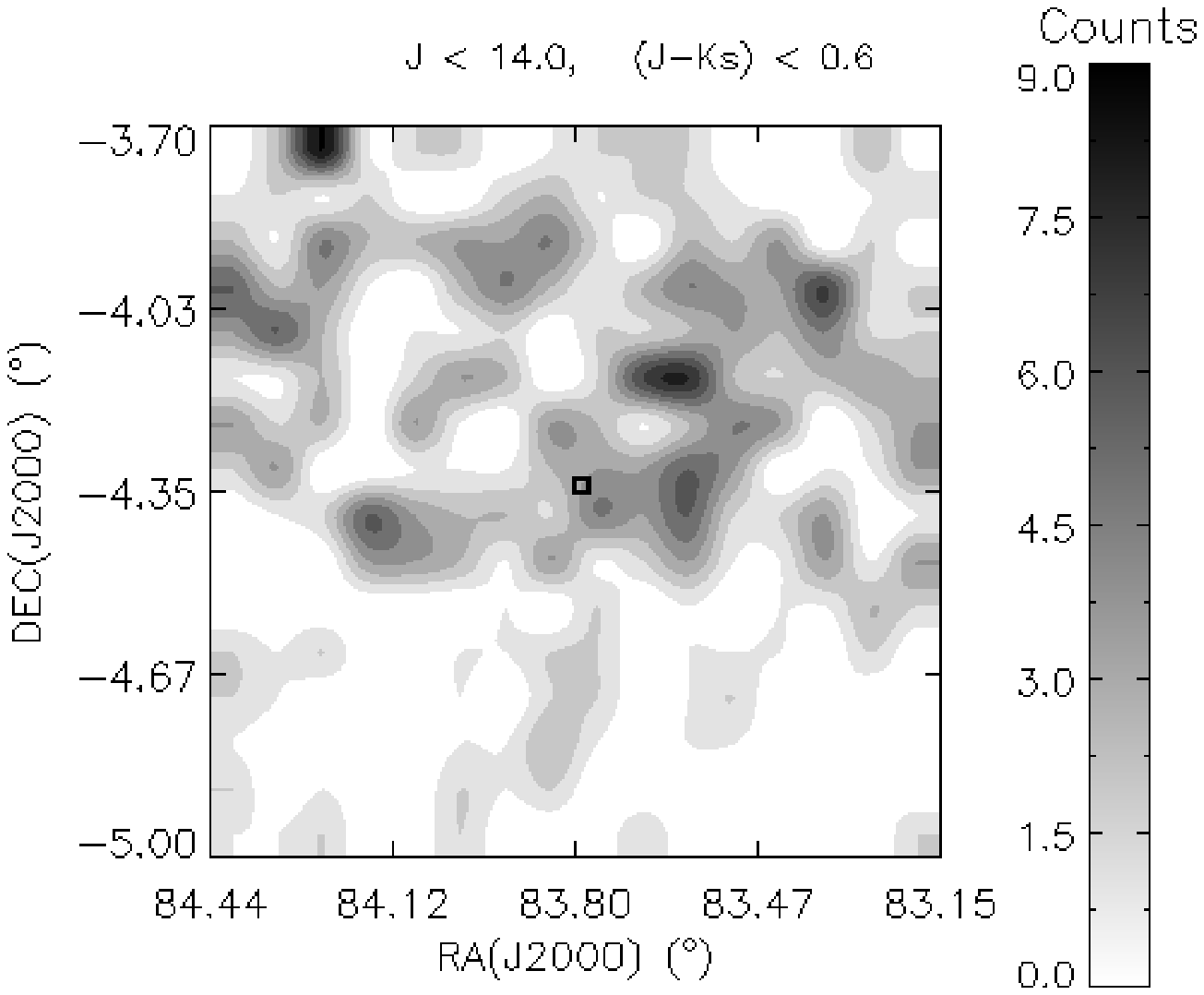}\\
\hskip -0.5cm
\includegraphics[width=0.5\linewidth]{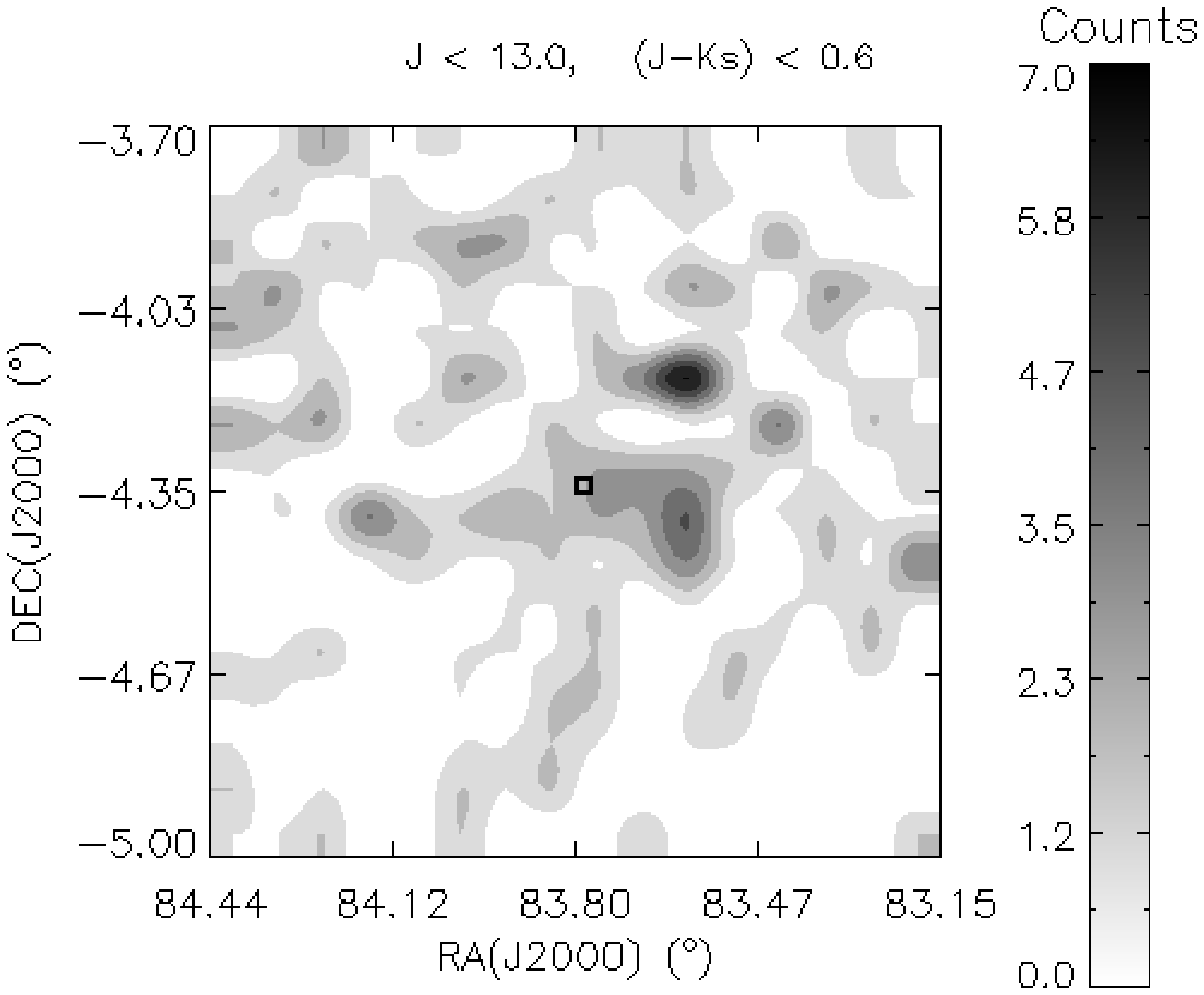}
\includegraphics[width=0.5\linewidth]{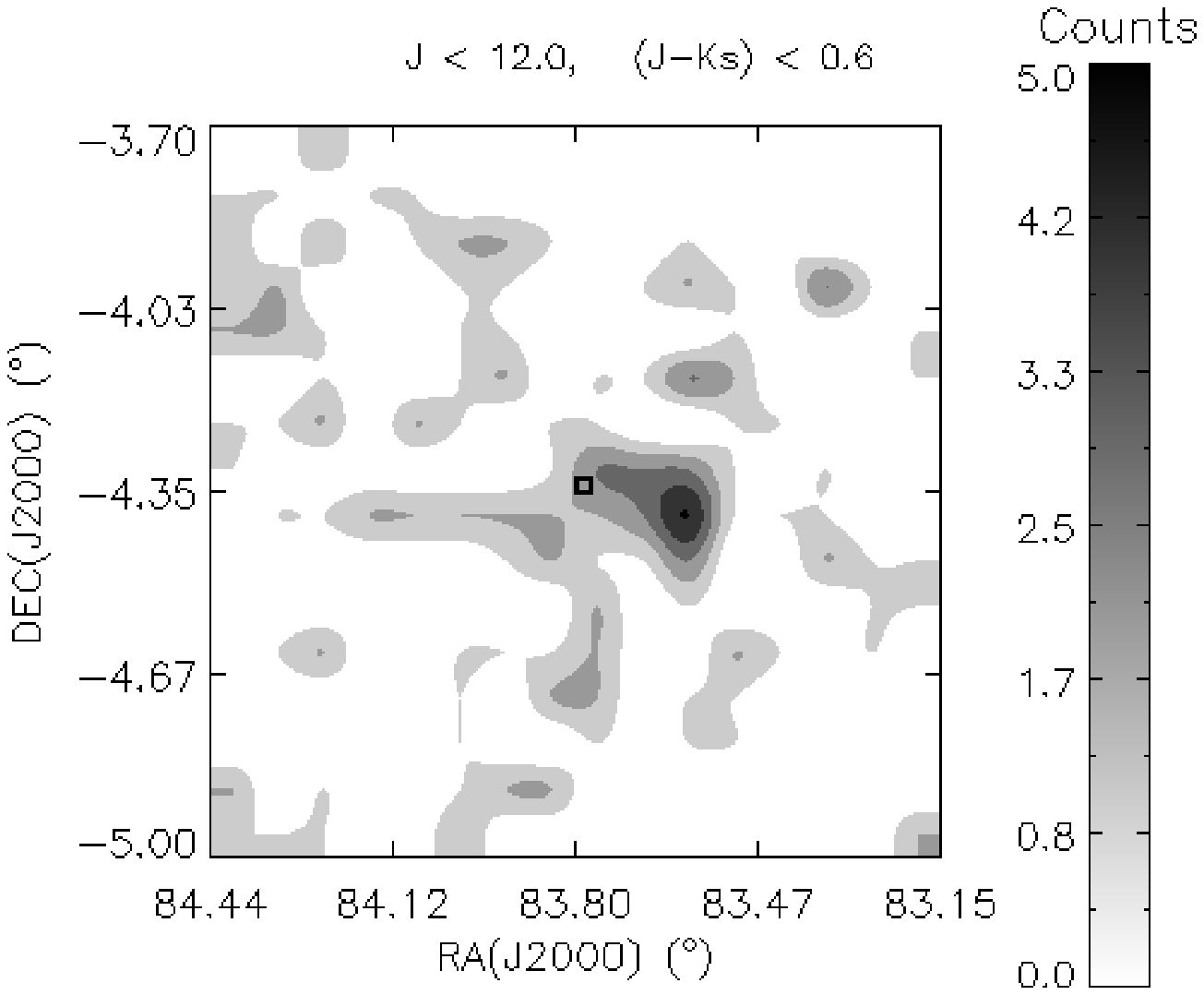}\\
\hskip -0.5cm
\includegraphics[width=0.5\linewidth]{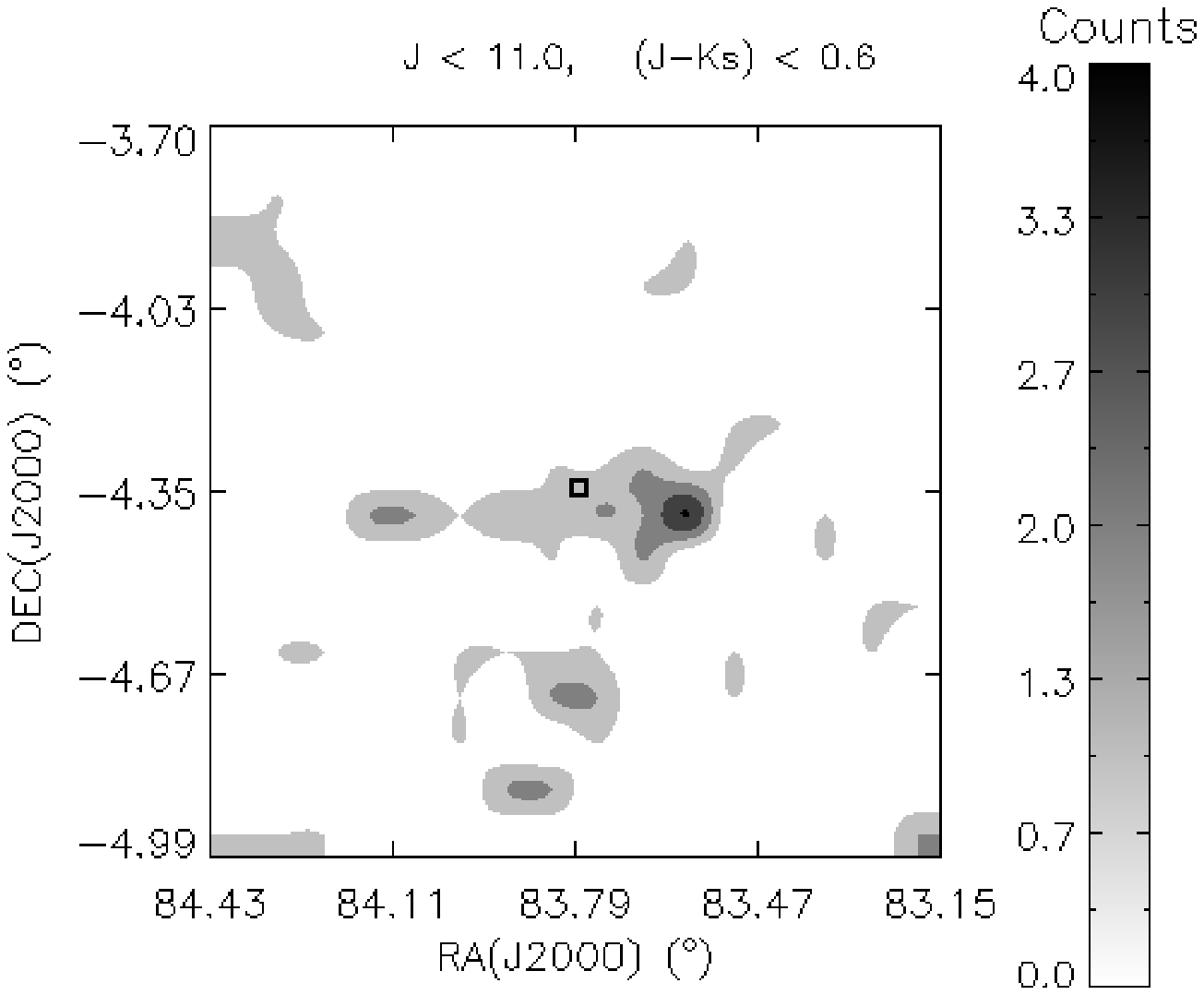}
\includegraphics[width=0.5\linewidth]{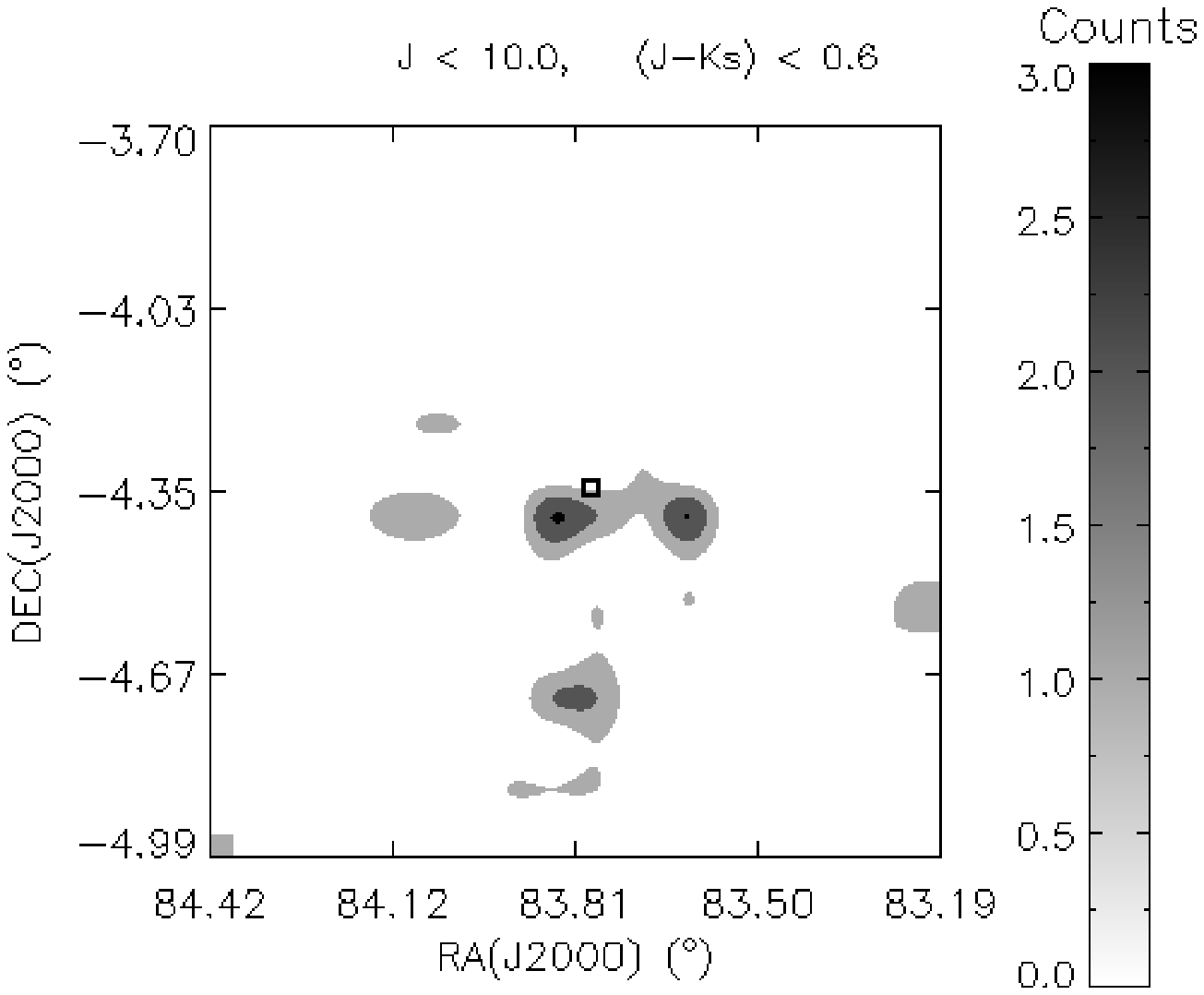}
\caption{Star density maps for ($J-K_s$)$<0.6$ and variable
$J$ magnitude limit as indicated.}
\label{f:densmap}
\end{figure}

Fig. \ref{f:densmap} shows the spatial stellar density plots for the same
area as in Fig. \ref{f:denspart}, but with progressive cuts in star
brightness, allowing one to
connect the density peaks with different populations characterized by
a limit magnitude. The plots show the emergence of NGC\,1981 as the
limiting magnitude varies from $J = 15.8$ to 10.

Although NGC\,1981 is in a region of intense star formation with clouds
of dust and gas, it has presently inhabited a relatively dust-free field,
perhaps as a consequence of the cluster evolution. Its massive
stellar content may have contributed to the energy release into the
interstellar medium, either by means of exploding supernovae and/or
winds due to radiation pressure. Our study concludes that
NGC\,1981 is a cluster older than the ONC and NGC\,1977, which are still 
embedded in the parental gas and dust cloud, and that its structure is best 
visualized by considering
stars brighter than $J = 12$.

\begin{figure}
\centering
\hskip -0.5cm
\includegraphics[width=0.5\linewidth]{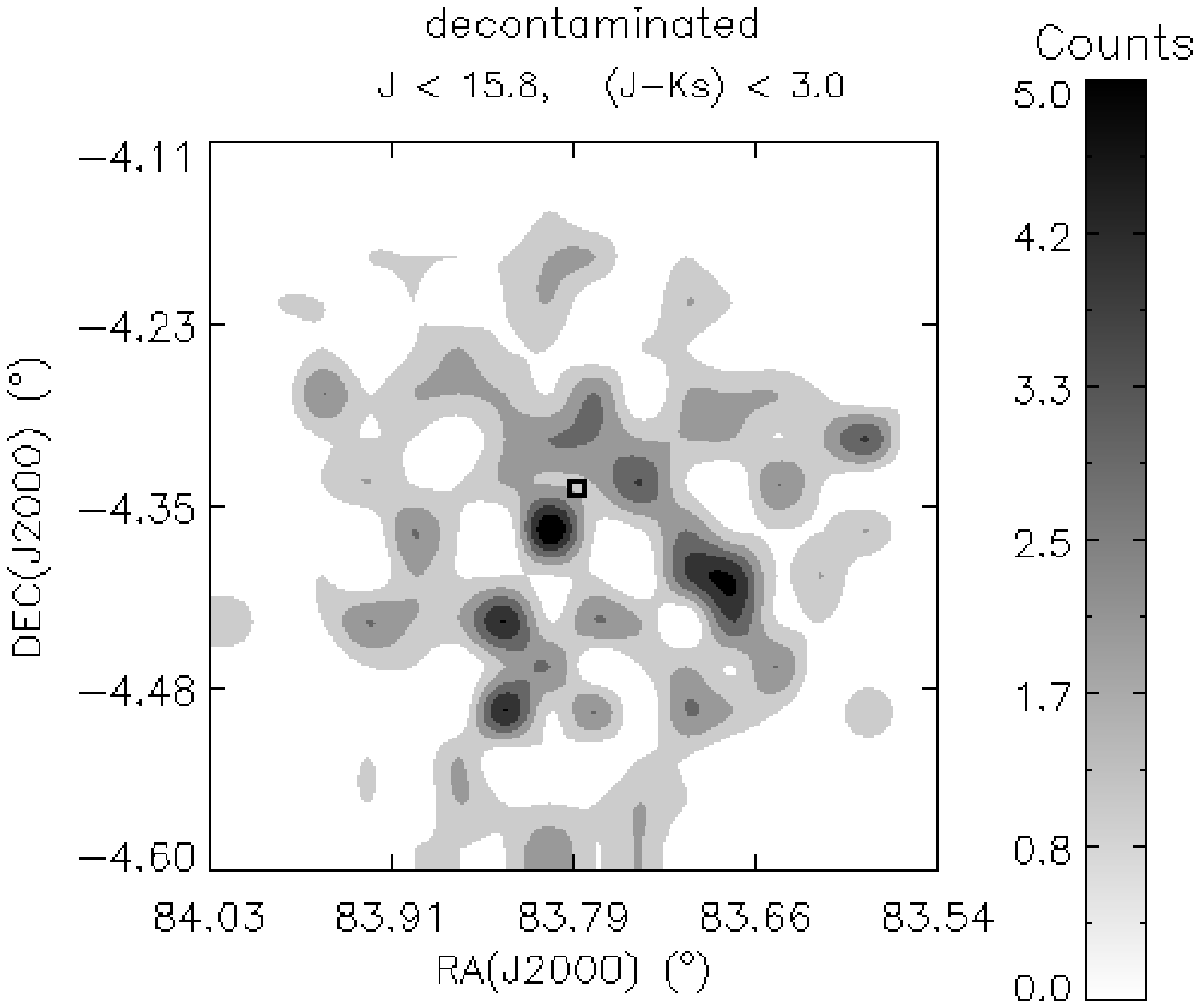}
\includegraphics[width=0.5\linewidth]{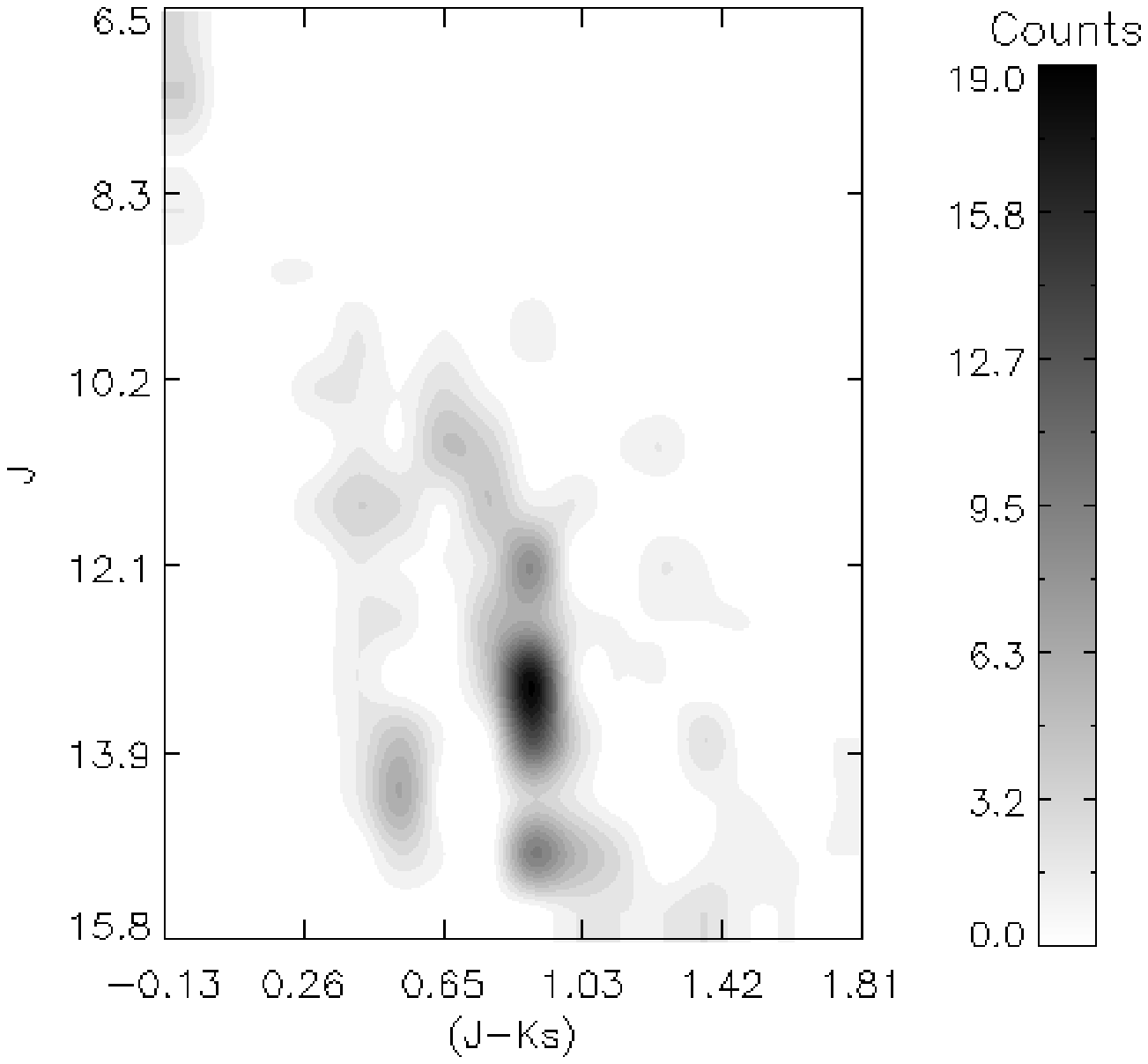}
\caption{Decontaminated star density map (left) and Hess diagram (right)
for a circular region of $R = 15$ arcmin centred in NGC\,1981.}
\label{f:mapdecont}
\end{figure}

Fig. \ref{f:mapdecont} shows the same analysis applied to probable members 
inside a circle of radius $R = 15$ arcmin centred in the cluster.
As a consequence of the method applied for field decontamination, the
cluster stellar content is more clearly defined down to the photometric
limit of the data. The underlying stellar field and the cluster stellar
population were indistinguishable in the previous contaminated diagrams
for stars fainter than $J = 12$ (Figs. \ref{f:densall}
to \ref{f:densmap}). In Fig. \ref{f:mapdecont}, a population of pre-main
sequence members stands out in the Hess diagram. 

\section{Discussion and Conclusions}

We investigated the stellar object NGC\,1981, a young star cluster in the 
Gould's Belt nearby the Orion Nebula Cluster containing a small population of 
massive B type stars. We employed near-infrared data from 2MASS 
to calculate the cluster centre at 
$\alpha=5^\mathrm{h} 35^\mathrm{m} 08^\mathrm{s}$, 
$\delta=-04^\circ 20\arcmin35\arcsec$ and to subsequently derive the structural 
parameters from King-profile fittings obtaining a core radius 
$R_\mathrm{c}=0.09 \pm 
0.04$ pc and central density $\sigma_0 = (2\pm1)\times 10^2$ stars.pc$^{-2}$ . 
A limiting radius of $R_\mathrm{lim}=1.21 \pm 0.11$ pc was also derived from 
the radial density profile. 

We devised a decontamination procedure based on the method by \citet{bb07} to 
statistically remove the underlying field population from the cluster CMD by 
using an offset-field to sample the background contamination. 
Tested on NGC\,1981 with
multiple offset-fields, the procedure reliably presented an average field-star 
exclusion efficiency of 84 per cent, separating cluster members with an average 
deviation of 5 per cent in the number of stars. It also provides photometric 
membership for the member stars with average uncertainty of 6 per cent, 
depending on the selection of the offset-field.

By using optical $BV(RI)_C$ data alongside 2MASS data we performed  
isochrone 
fittings on the decontaminated data and determined average values of 
reddening $E(B-V)=0.07\pm 0.03$, distance modulus $(m-M)=7.9\pm0.1$ 
($d=380\pm 17$ pc) and age $5\pm1$ Myr. 
The relation between the turn-on $M_V$ and age \citep{Cignoni2010} was also 
used as an additional check for the cluster age. In this case the age derived
was $3.3\pm2.7$ Myr.
The scarcity of bright stars make this value unreliable as sole 
indicator of age. We use it to infer that the luminosity of the 
turn-on is consistent with the age and distance modulus found by isochrone 
fittings.

By comparing our derived memberships with proper motions from UCAC2 
catalogue, we showed that while the small statistical distance between 
cluster and field populations may have hardened their separation 
in the VPD, the photometric method was capable to discern 
these very entangled populations. Indeed, by comparing our derived memberships
with proper motions from UCAC3, we were able to distinguish a clear 
concentration of member stars over a sparse distribution of non-member stars 
in the VPD.

We derived mass functions for stars inside the inner 5.5 arcmin and 
11 arcmin (limiting radius) and evaluated the total mass inside these regions: 
$m =  30 \pm 6$ M$_\odot$ and $m =  107 \pm 13$ M$_\odot$, respectively.
After fitting a power law to the data, the calculated slopes $\chi = 0.65\pm 
0.08$ (inner region) and $\chi = -0.44\pm 0.03$ (outer region) are flatter than
the Salpeter MF, indicating a depletion of low mass stars in the outer region 
relative to the inner region. 
Cluster evolution and mass segregation can explain this effect if the cluster 
has had time to dynamically evolve. 
In fact, by using proper motion data from DALM02 for stars inside the
limiting radius we have derived $t_\mathrm{cross}=0.15$ Myr and 
$t_\mathrm{relax}=0.41$ Myr, meaning that the 
cluster stars had time to interact gravitationally with 
each other forming binaries and/or multiple systems composed by massive stars 
that settle in the cluster centre and leading to the ejection
of low mass stars towards the cluster outer regions. 

The cluster might have undergone mass loss 
in the earlier stages of the star forming process leading to the present 
sparse structure. The energy released into the medium by the evolution of the
most massive stars (i.e. by supernovae and stellar winds) may have cleared the 
cluster from its parental cloud of gas and dust causing a collapse of the core 
due to the changing gravitational potential and subsequent evaporation of stars
in its outer regions. 

Through stellar density maps we discriminated the presence of a embedded red 
population of stars just south of NGC\,1981. These stars are likely members 
of the young object NGC\,1977 and still contaminate the southern population 
of NGC\,1981. These maps also demonstrate the power of the procedure employed 
to disentangle field-stars from the cluster population.

As the majority of young open 
clusters, NGC\,1981 is not expected to live longer than a few Myr, evolving 
from its actual state of marginally bound system to a loose stellar 
association and finally dispersing itself into the Galactic disc.

\section*{Acknowledgments}
We thank the referee, G. Carraro, for helping to improve this paper. We also 
thank C. Bonatto for the insightful comments that helped to develop this work.
We thank the Brazilian financial agencies FAPEMIG (grants APQ00154/08,  
APQ00117/08) and CNPq. We also thank the OPD staff for their support at the 
observatory. This publication makes use of data
products from the Two Micron All Sky Survey, which is a joint project of the
University of Massachusetts and the Infrared Processing and Analysis
Center/California Institute of Technology, funded by the National Aeronautics
and Space Administration and the National Science Foundation.
This research has made use of the WEBDA database, operated at the Institute
for Astronomy of the University of Vienna, and of the SIMBAD database,
operated at CDS, Strasbourg, France. This research has made use of Aladin.

\label{lastpage}


\begin{thebibliography}{99}

\bibitem[\protect\citeauthoryear{Bally}{2008}]{Bally2008} 
Bally J., 2008, in Bo Reipurth, ed., ASP Monograph 4, Handbook of Star Forming 
Regions, Vol I: The Northern Sky, p. 459

\bibitem[\protect\citeauthoryear{Binney \& Tremaine}{1987}]{Binney}
Binney J., Tremaine S., 1987, Galactic Dynamics, Princeton University Press, 
Princeton, NJ 

\bibitem[\protect\citeauthoryear{Bonatto, Bica \& Girardi}{Bonatto et al.}{2004}]{bbg04} 
Bonatto C., Bica E. Girardi L., 2004, A\&A, 415, 571

\bibitem[\protect\citeauthoryear{Bonatto, Santos Jr. \& Bica}{Bonatto et al.}{2006}]{bsb06} 
Bonatto C., Santos Jr. J.F.C., Bica E., 2006, A\&A, 445, 567

\bibitem[\protect\citeauthoryear{Bonatto \& Bica}{2007}]{bb07} 
Bonatto C., Bica E., 2007, MNRAS, 377, 1301

\bibitem[\protect\citeauthoryear{Cabrera-Ca\~no \& Alfaro}{1990}]{Cabrera90} 
Cabrera-Ca\~no J., Alfaro E.~J., 1990, A\&A, 235, 94 

\bibitem[\protect\citeauthoryear{Cignoni et al.}{2010}]{Cignoni2010} 
Cignoni M., Tosi M., Sabbi E., Nota A., Degl'Innocenti S., Prada Moroni P.G.,
Gallagher J.S., 2010, ApJ, 712, L63 

\bibitem[\protect\citeauthoryear{D'Antona}{2002}]{d02} 
D'Antona F., 2002, IAUS, 207, 599

\bibitem[\protect\citeauthoryear{Dias et al.}{2002}]{DAML02} 
Dias W.S., Alessi B.S., Moitinho A., Lepine J.R.D., 
2002, A\&A, 389, 871

\bibitem[\protect\citeauthoryear{Dias et al.}{2006}]{Dias06} 
Dias W.S., Assafin M., Fl{\'o}rio V., Alessi B.S., L\'{\i}bero V.,
2006, A\&A, 446, 949 

\bibitem[\protect\citeauthoryear{Elias, Alfaro \& Cabrera-Ca{\~n}o}{Elias et al.}{2009}]{eac09} 
Elias F., Alfaro E.J., Cabrera-Ca{\~n}o J., 2009, MNRAS, 397, 2 

\bibitem[\protect\citeauthoryear{Hron}{1987}]{h87} 
Hron J., 1987, A\&A, 176, 34

\bibitem[\protect\citeauthoryear{Jacobson, Pilachowski \& Friel}{Jacobson et al.}{2008}]{Jacobson} 
Jacobson H.R., Pilachowski C.A., Friel E.D., 2008, in J.G. Funes, E.M. Corsini,
eds, ASP Conf. Ser. Vol 396, Formation and Evolution of Galaxy Disks, p. 73 

\bibitem[\protect\citeauthoryear{Kharchenko et al.}{2005}]{Kharchenko2005} 
Kharchenko N.V., Piskunov A.E., R{\"o}ser S., Schilbach E., Scholz R.D., 
2005, A\&A, 438, 1163 

\bibitem[\protect\citeauthoryear{King}{1962}]{King1962} 
King I., 1962, AJ, 67, 471 

\bibitem[\protect\citeauthoryear{Lada \& Lada}{2003}]{Lada03} 
Lada C.J., Lada E.A., 2003, ARA\&A, 41, 57 

\bibitem[\protect\citeauthoryear{Landin et al.}{1996}]{lvd96}	
Landin N.R., Ventura P., D'Antona F., Mendes L.T.S., Vaz L.P.R.,
2006, A\&A, 456, 269	

\bibitem[\protect\citeauthoryear{Landolt}{1992}]{Landolt} 
Landolt A.U., 1992, AJ, 104, 340 

\bibitem[\protect\citeauthoryear{Lesh}{1968}]{l68} 
Lesh J.R., 1968, ApJS, 17, 371

\bibitem[\protect\citeauthoryear{Maia}{2007}]{Maia}
Maia F.F.S., 2007, Master Dissertation, UFMG

\bibitem[\protect\citeauthoryear{Marigo et al.}{2008}]{Marigo08} 
Marigo P., Girardi L., Bressan A., Groenewegen M.A.T., Silva L., Granato G.L.,
2008, A\&A, 482, 883 

\bibitem[\protect\citeauthoryear{Muench et al.}{2008}]{Muench} 
Muench A., Getman K., Hillenbrand L., Preibisch T. 2008, in Bo Reipurth, ed.,
ASP Monograph 4, Handbook of Star Forming Regions, Vol I: The Northern Sky,
p. 483

\bibitem[\protect\citeauthoryear{Munari \& Carraro}{1996}]{Munari96} 
Munari U., Carraro G., 1996, A\&A, 314, 108

\bibitem[\protect\citeauthoryear{Pavani \& Bica}{2007}]{Pavani2007} 
Pavani D.B., Bica, E., 2007, A\&A, 468, 139 

\bibitem[\protect\citeauthoryear{Piatti, Clari\'a \& Abadi}{Piatti et al.}{1995}]{pca95} 
Piatti A.E., Clari\'a J.J., Abadi M.G., 1995, AJ, 110, 2813	

\bibitem[\protect\citeauthoryear{Piskunov et al.}{2006}]{Piskunov} 
Piskunov A.E., Kharchenko N.V., Röser S., Schilbach E., Scholz R.D.,
2006, A\&A, 445, 545

\bibitem[\protect\citeauthoryear{P\"oppel}{2001}]{p01} 
P\"oppel, W.G.L. 2001, in T. Montmerle \& P. Andr\'e, eds, ASP Conf. Ser. Vol 
243, From Darkness to Light: Origin and Evolution of Young Stellar Clusters, 
Astron. Soc. Pac., San Francisco, p. 667

\bibitem[\protect\citeauthoryear{Rieke \& Lebofsky}{1985}]{Rieke85} 
Rieke G.H., Lebofsky M.J., 1985, ApJ, 288, 618 

\bibitem[\protect\citeauthoryear{S{\'a}nchez \& Alfaro}{2009}]{Sanchez09} 
S{\'a}nchez N., Alfaro E.~J., 2009, ApJ, 696, 2086 

\bibitem[\protect\citeauthoryear{Salpeter}{1955}]{Salpeter}
Salpeter E., 1955, ApJ, 121, 161

\bibitem[\protect\citeauthoryear{Santos Jr., Bonatto \& Bica}{Santos Jr. et al.}{2005}]{sbb05} 
Santos Jr. J.F.C., Bonatto C., Bica E., 2005, A\&A, 442, 201

\bibitem[\protect\citeauthoryear{Schlegel, Finkbeiner \& Davis}{Schlegel et al.}{1998}]{sfd98} 
Schlegel D.J., Finkbeiner D.P., Davis M., 1998, ApJ, 500, 525 

\bibitem[\protect\citeauthoryear{Schmidt-Kaler}{1982}]{SK}
Schmidt-Kaler Th., 1982, in Landolt-Bornstein New Serires Vol. 2b: Astronomy
and Astrophysics/Star and Star clusters. Springer-Verlag, Newbak, p. 14 

\bibitem[\protect\citeauthoryear{Sestito et al.}{2008}]{Sestito} 
Sestito P., Bragaglia A., Randich S., Pallavicini R., 
Andrievsky S.M., Korotin S.A., 2008, A\&A, 488, 943 

\bibitem[\protect\citeauthoryear{Sharpless}{1952}]{sha52} 
Sharpless S., 1952, ApJ, 116, 251

\bibitem[\protect\citeauthoryear{Siess, Dufour \& Forestini}{Siess et al.}{2000}]{Siess2000} 
Siess L., Dufour E., Forestini M., 2000, A\&A, 358, 593 

\bibitem[\protect\citeauthoryear{Skrutskie et al.}{2006}]{2mass} 
Skrutskie M.F., Cutri R.M., Stiening R., et al., 2006, AJ, 131, 1163

\bibitem[\protect\citeauthoryear{Subramaniam et al.}{1995}]{sgsb95} 
Subramaniam A., Gorti U., Sagar R., Bhatt H.C., 1995, A\&A, 302, 86

\bibitem[\protect\citeauthoryear{Zacharias et al.}{2004}]{ucac2} 
Zacharias N., Urban S.~E., Zacharias 
M.~I., Wycoff G.~L., Hall D.~M., Monet D.~G., Rafferty T.~J., 2004, AJ, 
127, 3043

\bibitem[\protect\citeauthoryear{Zacharias et al.}{2010}]{ucac3} 
Zacharias N., et al., 2010, AJ, 139, 2184 

\end{thebibliography}
\end{document}